\documentclass[reprint, onecolumn, tightenlines, notitlepage, footinbib, 12pt, longbibliography]{revtex4-1}
\usepackage{mathrsfs}  
\usepackage{amsfonts}
\usepackage{amsmath}
\usepackage{amssymb}
\usepackage{epsfig}
\usepackage{natbib}

\usepackage{color}

\newcommand{\rarb}{\rayleigh_{\mathrm{RB}}}
\newcommand{\mri}{\mathrm{i}}

\newcommand{\Ek}{\mathrm{E}}

\newcommand{\nusselt}{\mathrm{Nu}}
\newcommand{\rayleigh}{\mathrm{Ra}}

\newcommand{\prandtl}{\mathrm{Pr}}
\newcommand{\stdof}[1]{\sigma_{#1}}
\newcommand{\ccite}[1]{\cite{#1}}
\newcommand{\EBC}{E$\beta$C~}
\newcommand{\D}{\mathrm{D}}
\newcommand{\integrate}[1]{ \int{\mathrm{d}^3 \mathcal{V} \left[  #1  \right] } }
\newcommand{\approxsol}{_{\mathrm{a.s.}}}
\newcommand{\kperp}{K_\perp^2}

\begin{document}
\title{Equatorially Trapped Convection in a Rapidly Rotating Shallow Shell}
\author{Benjamin Miquel}
\affiliation{Department of Applied Mathematics, University of Colorado, Boulder, CO 80309, USA}
\author{Jin-Han Xie}
\affiliation{Courant Institute of Mathematical Sciences, New York University, New York, NY 10012, USA}
\author{Nicholas Featherstone}
\affiliation{Department of Applied Mathematics, University of Colorado, Boulder, CO 80309, USA}
\author{Keith Julien}
\affiliation{Department of Applied Mathematics, University of Colorado, Boulder, CO 80309, USA}
\author{Edgar Knobloch}
\affiliation{Department of Physics, University of California, Berkeley, CA 94720, USA}
\date{21 April 2018}
\begin{abstract}
  Motivated by the recent discovery of subsurface oceans on planetary moons and the interest they have generated, we explore convective flows in shallow spherical shells of dimensionless gap width $\varepsilon^2\ll 1$ in the rapid rotation limit $\Ek\ll1$, where $\Ek$ is the Ekman number. We employ direct numerical simulations (DNS) of the Boussinesq equations to compute the local heat flux $\nusselt(\lambda)$ as a function of the latitude $\lambda$ and use the results to characterize the trapping of convection at low latitudes, around the equator. We show that these results are quantitatively reproduced by an asymptotically exact nonhydrostatic equatorial $\beta$-plane convection model at a much more modest computational cost than DNS. We identify the trapping parameter $\beta=\varepsilon \Ek^{-1}$ as the key parameter that controls the vigor and latitudinal extent of convection for moderate thermal forcing when $\Ek\sim\varepsilon$ and $\varepsilon\downarrow 0$. This model provides a new theoretical paradigm for nonlinear investigations.

\end{abstract}
\maketitle
%
%
%
%
\section{Introduction}
Recent exploration campaigns of the solar system have provided evidence of the presence of shallow subsurface global oceans of water in ice-clad moons of giant gaseous planets (in Europa~\ccite{europaNAT98,europaSCI00}, Titan~\ccite{titanSCI08}, Ganymede~\ccite{ganymedeSCI01}, Callisto~\ccite{europaNAT98}, Enceladus~\ccite{enceladusGRL15, enceladusICARUS16} and Triton~\ccite{tritonICARUS12}) or at the other extreme molten magma (in Io~\ccite{ioSCI11}). The convective flows in these oceans are believed to have far-reaching consequences for each moon including the active shaping of the outer crust topography~\ccite{europaICARUS04} and magnetic field generation via the dynamo effect~\ccite{europaNAT98}. For instance, the latitudinal modulation of heat flux carried by convection has been proposed to explain the chaotic topography of Europa's icy surface at low latitudes~\ccite{soderlundNAT14}.

However, convection in the rapid rotation regimes that generally pertain to planets or their moons remains an arduous problem. Such regimes are characterized mathematically by geostrophic balance between the Coriolis force and pressure gradient, resulting in the Proudman-Taylor constraint that gives rise to the formation of highly anisotropic Taylor columns aligned with the rotation axis~\ccite{robertsPTRSA68,busseJFM70,yanoJFM92,zhangJFM92, dormyJFM04,bussePOF02}. This anisotropy and the multiple scales associated with it have prevented theoretical or numerical investigations of the nonlinear regimes of convection approaching the strongly rotationally constrained regimes~\cite{aurnouPEPI15}. The shallow aspect ratio geometry (studied experimentally in a microgravity environment in~\ccite{hartJFM86}) compounds the difficulty and hinders direct numerical integration of the governing equations. 

One possible strategy that has proved successful in related problems is asymptotic model reduction. In this approach the system parameters are linked in a careful and precise way, and an asymptotic expansion is performed that is valid in the limit in which this parameter approaches zero or infinity. The resulting leading order equations provide an asymptotically exact reduction of the primitive equations provided they are closed. These techniques have proved remarkably effective in establishing reliable reduced (i.e., simplified) model equations for a variety of buoyancy-driven geophysical flows that are valid in extreme but geophysically or astrophysically relevant parameter regimes. These include rapidly rotating convection in the polar regions of rotating shells~\cite{spragueJFM06} and studies of its potential for dynamo action~\cite{calkinsJFM15}, following the seminal contribution of~\cite{childressPRL72}, and turbulent doubly diffusive convection in both oceans and stars in the fingering regime~\cite{xieFLUIDS17}, to cite two recent examples. In both these cases the domains invoked are simple: a three-dimensional horizontal layer with periodic boundary conditions in the horizontal in the former, and a triply-periodic domain in the latter.

\begin{figure}
\includegraphics[width=0.6\columnwidth]{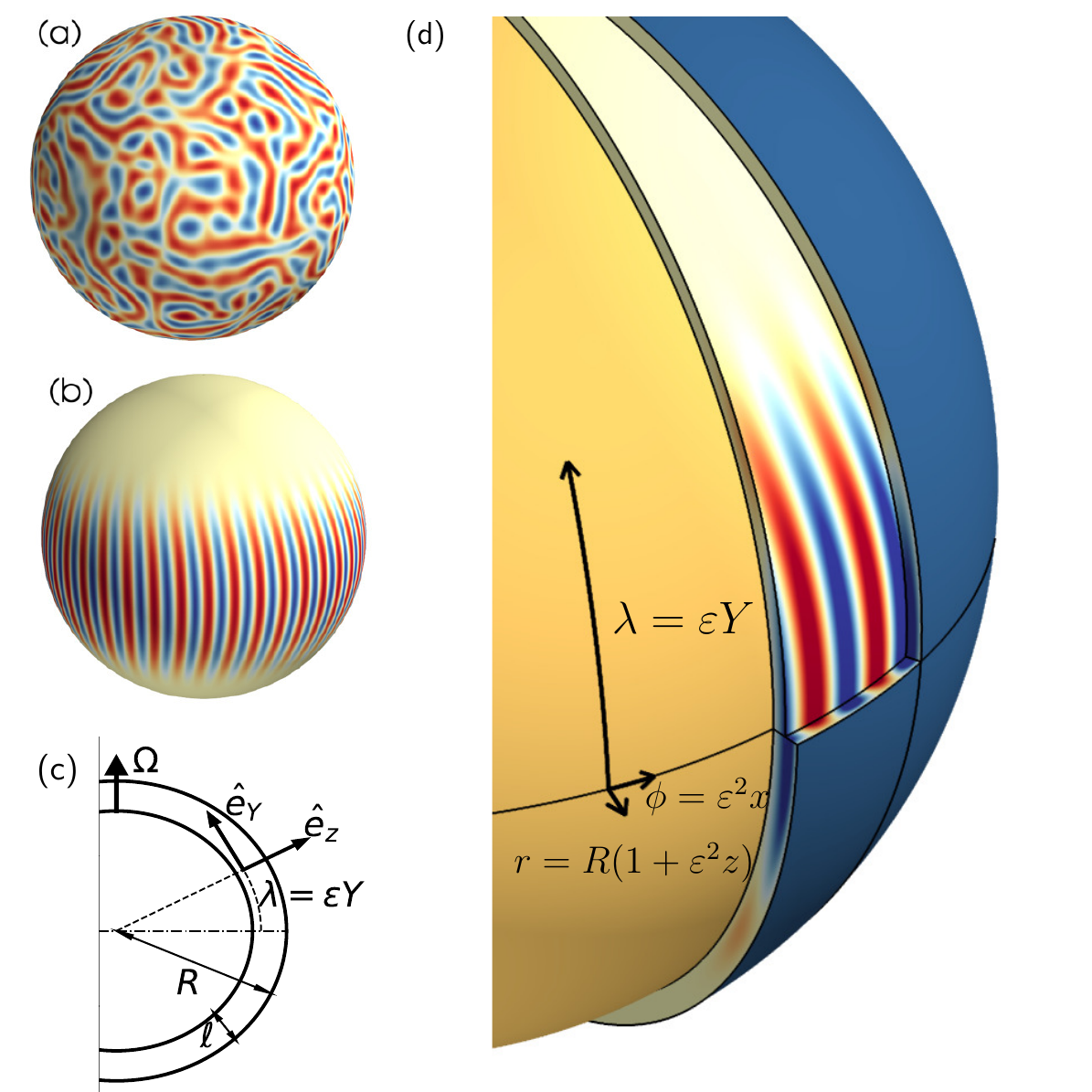}
\caption{Shell slices of temperature taken at mid-depth of a shallow shell with aspect ratio $\ell/R = 0.05$ in (a) slow and (b) rapid rotation. Equatorial and meridional slices of the temperature field in case (b) are represented in a 3D visualization in panel (d). In (b) and (d) convective motions adopt the form of anisotropic Taylor columns localized in the equatorial region. The equatorial reduced coordinates $(x,Y,z)$ [see Eq.~(\ref{anisotropic_scaling})] are represented from the side in panel (c) and in perspective in panel (d).\label{fig_intro}}
\end{figure}

We follow here the above approach and introduce an asymptotically reduced model for convection in the equatorial $\beta$-plane by leveraging the following idea: in a shallow rotating shell, elongated Taylor columns develop preferentially in the vicinity of the equator, where the tangent plane is locally aligned with the rotation axis (Fig.~\ref{fig_intro}). The resulting \emph{nonhydrostatic} equatorial $\beta$-plane convection (E$\beta$C) model, a nonlinear counterpart of the classical (\emph{hydrostatic}) equatorial $\beta$-plane model for atmospheric waves~\ccite{gillBOOK}, successfully captures the nonlinear dynamics of equatorial convective flows in shallow shells and provides predictions for the latitude dependence of the heat flux pertaining to global modelling of the oceans and solid-state convection in ice crusts of both planets and their moons. 

Prior to a rigorous presentation of the \EBC model, we wish to  provide some physical intuition behind the development of this model. The \EBC model is akin to a plane Rayleigh-B\'enard setup with rotation about an axis $Y$ (playing the role of latitude) parallel to the top or bottom boundary. However, the model includes the variation of the rotation vector with latitude via a normal component of rotation vector that varies linearly with $Y$ and therefore vanishes at the equator $Y=0$ where it changes sign. All other contributions of sphericity represent higher order effects and so are absent from the \EBC model.

The paper is organized as follows. The governing equations for fluid shells and the resulting \EBC model are presented in Sec.~\ref{sec:formulation}, along with a discussion of the trapping mechanism that is responsible for the confinement of the convection cells in the equatorial region. Throughout the paper we compare the solutions of the \EBC model with those of the primitive equations in order to validate the model. The numerical methods used for this purpose are described in Sec.~\ref{sec:methods}. An analysis of the linear stage of the convective instability appears in Sec.~\ref{sec:linear} while the resulting saturated state and the corresponding equatorially-trapped heat flux are characterized in Sec.~\ref{sec:nonlinear}. Finally, closing remarks are found in Sec.~\ref{sec:conclusion}.
%
%
%
%
\section{Formulation}
\label{sec:formulation}
\subsection{Governing equations {\color{red}and dimensionless parameters}}
For simplicity, we consider a Boussinesq fluid with kinematic viscosity $\nu$, thermal conductivity $\kappa$ and thermal expansion coefficient $\alpha$ contained within a spherical shell of outer radius $R$ and gap $\ell\ll R$ rotating around an axis $\boldsymbol{e}_\Omega$ with a constant angular velocity $\Omega$ [see Fig.~\ref{fig_intro}(c)]. A constant temperature difference $\delta\bar{\Theta}$ is imposed between the two bounding surfaces and is used as a temperature unit below. Gravity is radial, directed inwards, with strength $gr^*/R$ proportional to the dimensional radial coordinate $r^*$, where $g$ represents the gravitational acceleration at the outer shell. In the absence of motion ($\overline{\boldsymbol{u}}=0$), the dimensionless background temperature profile is spherically symmetric: $\overline{\Theta}(r^*)  = \overline{\Theta}_0 +  R\left(R-\ell\right) /(\ell r^*)$. The dimensionless fluctuations of the velocity $\boldsymbol{u}$ and temperature $\Theta$ obey the Navier-Stokes equations (NSE):
\begin{subequations}\label{NSE}
\begin{gather}
\prandtl^{-1} \left(\partial_t\boldsymbol{u} + \boldsymbol{u\cdot\nabla u}\right)+ \Ek^{-1}\boldsymbol{e}_\Omega \wedge \boldsymbol{u} = - \Ek^{-1}\boldsymbol{\nabla}p  + \varepsilon^2 r\, \rayleigh\, \Theta\, \boldsymbol{e}_r + \nabla^2 \boldsymbol{u} \,,\label{NS_momentum} \\ 
\boldsymbol{\nabla\cdot u} =0 \,,\label{NS_div}\\
\partial_t\Theta + \boldsymbol{u\cdot \nabla} \left(\overline{\Theta} + \Theta\right)  = \nabla^2 \Theta\,,\label{NS_temp} %
\end{gather}
\end{subequations} 
where $r = r^*/\ell$ is the dimensionless radial coordinate. We define the Prandtl number $\prandtl$, the gap-based Ekman number $\Ek$, the Rayleigh number $\rayleigh$, and the dimensionless gap $\varepsilon^2$ as follows:
\begin{equation}
\prandtl = \frac{\nu}{\kappa},\qquad \Ek =\frac{\nu} {2\Omega \ell^2},\qquad \rayleigh = \frac{\alpha g \ell^3 \delta\bar{\Theta} }{\kappa\nu},\qquad \varepsilon^2 = \frac{\ell}{R}\,.
\end{equation}
In the following, motivated by the case of planetary moons (see Table~\ref{tab:moons_features} and the closing discussion in Sec.~\ref{sec:discussion}), we consider the limit of shallow shells $\varepsilon\ll 1$ in rapid rotation $\Ek \ll 1$, under the assumption that both the Prandtl and the Rayleigh numbers are $\mathcal{O}(1)$ quantities, and focus on the distinguished regime
\begin{equation}
\Ek \sim \varepsilon\,. \label{distinguished}
\end{equation}
 In this distinguished regime convective motions are localized around the equator, as shown next, and their latitudinal extent is controlled by the confinement parameter $\beta$, obtained by collapsing the two dimensionless parameters $\Ek$ and $\varepsilon$:
\begin{equation}
\beta = \varepsilon \Ek^{-1} = \mathcal{O}(1)\,.
\end{equation}  
The parameter $\beta$ quantifies the local rate of change of the Coriolis parameter at the equator: $\beta$ therefore increases with both the rotation rate (as measured by $\Ek$) and the gap $\varepsilon$. An alternative (and dual) view is that $\beta$ decreases with the local curvature of the shell.
\subsection{Asymptotic analysis and the reduced model} 
In this section we provide a concise sketch of the derivation of the Equatorial $\beta$-Convection (\EBC) model. The reader is referred to Appendix~\ref{appendix:derivation} for details of this derivation.

If buoyancy remains modest compared to the Coriolis force ($\rayleigh \ll \Ek^{-1}$), the leading order terms in Eq.~(\ref{NS_momentum}) yield the geostrophic balance $\boldsymbol{e}_\Omega \wedge \boldsymbol{u}_0=-\boldsymbol{\nabla}p_0$, responsible along with incompressibility for the Proudman-Taylor theorem invoked above: the flow tends to form anisotropic Taylor columns aligned with the rotation axis $\boldsymbol{e}_\Omega$ and localized around the equator. Variations along the rotation axis occur on a scale $\Ek^{-1}\ell$, i.e. on a scale larger by a factor $\Ek^{-1}$ than variations orthogonal to this axis. 
We therefore define an anisotropic set of local coordinates $(x,Y,z)$ around the equator:
\begin{equation}
\phi=\varepsilon^2x,\qquad \lambda = \varepsilon Y,\qquad r = \frac{1}{\varepsilon^2}(1+\varepsilon^2 z )\, , \label{anisotropic_scaling}
\end{equation}
where $\phi$ is the azimuthal angle and $\lambda$ is the latitude. The fluid is confined in the domain $z\in[-1,0]$. Note that the equatorial coordinates $x$ and $z$ correspond to a length set by the gap. In contrast, the reduced latitude $Y$ corresponds to an intermediate (geometric-mean) scale of order $\varepsilon R$, shorter than the radius of the sphere yet longer than the gap. This scale anisotropy, clearly depicted in Fig.~\ref{fig_intro}(d), is emphasized here by employing lower case symbols for $(x,z)$ and an upper case symbol for the latitude variable $Y$. Despite this anisotropy, the condition $Y=\mathcal{O}(1)$ defines a region confined to small latitudes.

Each variable $\boldsymbol{u}$, $p$ and $\Theta$ is now expanded in a series in $\varepsilon$, for instance:
\begin{equation}
\boldsymbol{u} = \boldsymbol{u}_0 + \varepsilon \boldsymbol{u}_1 + \varepsilon^2 \boldsymbol{u}_2+\dots\,. \label{power_expansion}
\end{equation}
The leading order geostrophic balance suggests that we decompose the velocity $\boldsymbol{u}_0$ into an equatorial stream function $\psi$ identical to the pressure, and a meridional velocity $V$, both of which are functions of $x$, $Y$ and $z$:
\begin{subequations}\label{leading_order}
\begin{gather}
\boldsymbol{u}_0(x,Y,z) = \partial_z\psi \,\boldsymbol{e}_x + V\, \boldsymbol{e}_y - \partial_x \psi\, \boldsymbol{e}_z\,,\\
p_0(x,Y,z) = \psi\,. 
\end{gather}
\end{subequations}
Evolution equations for $\psi$ and $V$ are obtained at the next order of the asymptotic hierarchy from a solvability condition for $\boldsymbol{u}_1$, $p_1$, $\psi$ and $V$ (see e.g.~\ccite{julienJFM06}). Relegating details to Appendix~\ref{appendix:derivation}, these equations and the corresponding temperature equation take the form:
\begin{subequations}\label{RME}
\begin{gather}\prandtl^{-1}D_t^\perp \nabla_\perp^2 \psi 
 - \beta\mathscr{M} V =  - \rayleigh\,\partial_x \Theta+ \nabla^4_\perp \psi  \,, \label{RM_psi}
 \\
 \prandtl^{-1}D_t^\perp V + \beta\mathscr{M} \psi=\nabla^2_\perp V\label{RM_V}\,,
  \\ 
 D_t^\perp \Theta+\partial_x \psi = \nabla_\perp^2\Theta\,,      \label{RM_theta}
 \end{gather}
\end{subequations}
where we have defined the equatorial diffusion operator $\nabla_\perp^2 = \partial_{xx} + \partial_{zz}$, the equatorial material derivative $D_t^\perp = \partial_t + \partial_x \psi \partial_z  - \partial_z \psi \partial_x$, and the meridional operator $\mathscr{M} = Y\partial_z + \partial_Y$. 

These equations are supplemented with appropriate boundary conditions at $z=-1$ (inner shell) and $z=0$ (outer shell). We consider either stress-free $\psi = \partial_{zz}\psi = \partial_z V=0$ or no-slip $\psi = \partial_z\psi = V=0$ velocity boundary conditions, together with fixed temperature $\Theta=0$ or fixed flux $\partial_z \Theta=0$ thermal boundary conditions. Periodic boundary conditions are natural in the azimuthal direction $x$. We also require boundedness of the solutions as $|Y| \rightarrow \infty$.

The asymptotically reduced Eqs.~(\ref{RME}), dubbed the \EBC model, are closely related to the usual equations describing two-dimensional Rayleigh-B\'enard convection (RBC) which are recovered by setting $V=0$, $\beta=0$~\ccite{chandrasekharBOOK}. The \EBC model with $\beta>0$ incorporates the latitudinal variation of the Coriolis parameter present in the equatorial region. This new effect is captured via the linear operator $\mathscr{M}$ which couples the equatorial stream function $\psi$ and the meridional velocity $V$. In the following section, we demonstrate that this coupling is responsible for the presence of equatorial trapping, i.e., the presence of localized convective motions in the form of ``banana cells''~\ccite{bussePOF02} around the equatorial plane of the shell.

  We remark that the breaking of the reflection symmetry $x\leftrightarrow -x$, a key feature of the primitive equations of motion in spherical geometries that leads to pattern precession, arises only at higher order in our asymptotic treatment. The \EBC model is thus invariant under the reflection $x\leftrightarrow -x$. As a result the slowly drifting modes usually found in rotating shells are absent from the \EBC model and the corresponding states do not precess. The implications of the absence of precession are discussed below.


%
\subsection{The quantum harmonic oscillator approximation}
\begin{figure}
\includegraphics[height = 0.6 \textwidth]{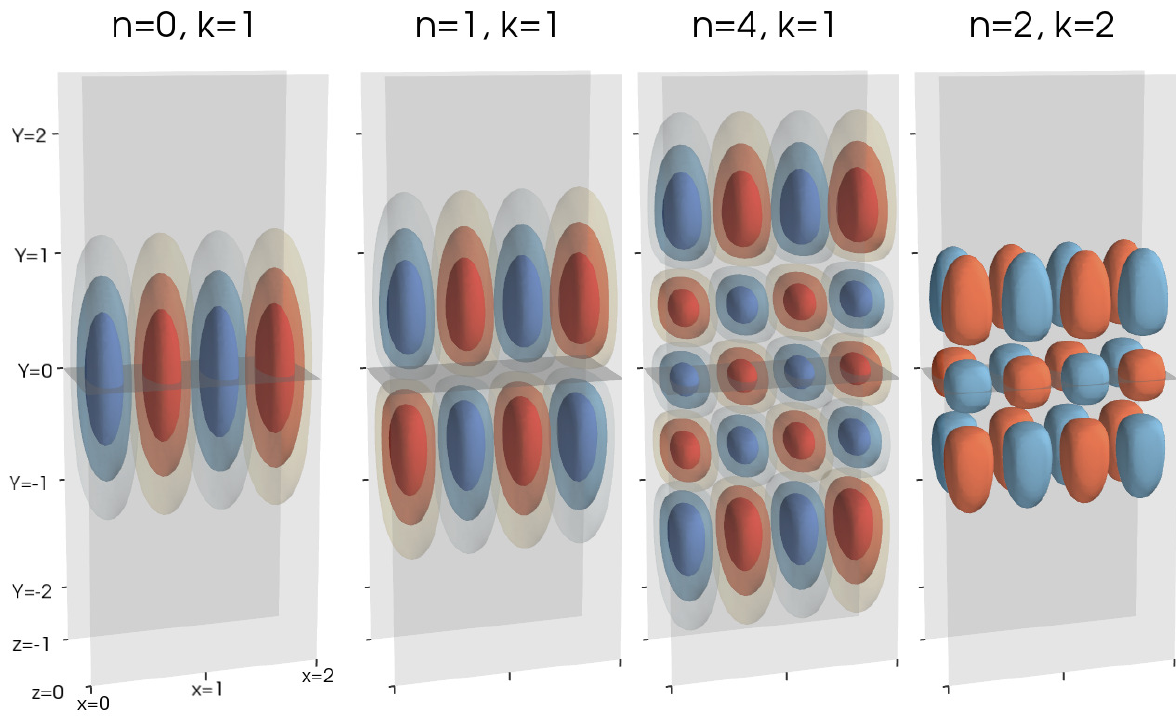}
\caption{\label{fig:banana_cells} Isosurfaces of the vorticity $\omega=\left(\partial_{xx}+\partial_{zz}\right)\psi\approxsol$ along the $Y$ axis for $n$-banana cells obtained from the quantum harmonic oscillator ansatz [Eqs.~(\ref{wild_ansatz}) and (\ref{eq:QHO_eigenfunction})] for different values of the parameters $k$ and $n$. The vorticity fields are normalized by their respective maximum value and isosurfaces are drawn for $\omega=\pm0.1$, $\pm0.3$ and $\pm0.6$. (For legibility, only the surfaces corresponding to $\omega=\pm0.3$ are represented when $k=2$.) The blue and red hues correspond to opposite cyclonicities. The equatorial plane at $Y=0$ is also indicated. }
\end{figure}
\label{sec:QHO_approx}
Before turning to a rigorous numerical study of Eq.~(\ref{RME}) to evidence the presence of banana cells, we present a semi-qualitative but analytic argument in favor of trapping as a result of the presence of the operator $\mathscr{M}$. In the linear limit, a stationary solution to Eq.~(\ref{RME}) obeys
\begin{equation}
-\beta^2\mathscr{M}^2 \psi = \left(-\rayleigh\, \partial_{xx} + \nabla_\perp^6\right)\psi\,,
\label{RRME}\end{equation}
and the variables $V$ and $\Theta$ follow by solving $\beta \mathscr{M}\psi = \nabla_\perp^2 V$, and $\partial_x\psi = \nabla^2_\perp\Theta$, respectively. Analytical solutions to Eq.~(\ref{RRME}) remain arduous to obtain owing to the complexity of the operator $\mathscr{M}^2 = \partial_{YY} + Y^2 \partial_{zz} + 2Y\partial_{Yz} + \partial_z$. Nevertheless, an attempt can be made at leveraging the kinship existing between Eq.~(\ref{RME}) and 2D RBC. In this spirit, we assume that, for stress-free boundary conditions, an approximate solution $\psi\approxsol$ to Eq.~(\ref{RRME}) has the separable form
\begin{equation}
\label{wild_ansatz}
\psi\approxsol \approx \sin(k\pi z)\exp(\mri m x)\Psi (Y)\,,
\end{equation}
where $k\in\mathbb{N^+}$ is a positive integer  and the variation along latitude is contained in the yet unknown function $\Psi(Y)$. This ansatz diagonalizes the right hand side of Eq.~(\ref{RRME}). Furthermore, the action of $\mathscr{M}^2$ simplifies considerably along mid-depth $z=-1/2$ because the $z$ derivative vanishes, and Eq.~(\ref{RRME}) then yields:
\begin{equation}
\label{QHO}
\left(-\partial_{YY} + k^2\pi^2 Y^2\right) \Psi(Y) = \frac{1}{\beta^2} \left(\rayleigh\,m^2 - \left[m^2+k^2\pi^2\right]^3\right) \Psi(Y)\,.
\end{equation} 
The left hand side of this equation corresponds to the Hamiltonian of a quantum particle in a harmonic well, whose well-known eigen-energies $E_n$ and eigenfunctions $\Psi_n(Y)$ are:
\begin{subequations}
\begin{gather}
\label{QHO_ev}
E_n = k\pi (1+2n)\,,\\
\Psi_n(Y) = H_n\left(\sqrt{k\pi}Y\right)\exp\left(-\frac{k\pi Y^2}{2}\right)\,, \label{eq:QHO_eigenfunction}
\end{gather}
\end{subequations}
where $n\in\mathbb{N}$ is an integer and $H_n(Y)$ is the Hermite polynomial of degree $n$ (see, e.g., \cite{cohen} or \cite{matsunoJMSJ66}). These eigenfunctions are depicted in Fig.~\ref{fig:banana_cells}. Thus, provided that a stationary solution exists in the linear limit and is reasonably described by our ansatz [Eq.~(\ref{wild_ansatz})], we can utilize Eq.~(\ref{QHO}) and solve for the onset of the mode $n$ along the $Y$ direction:
\begin{equation}
\rayleigh^{(QHO)}(m,n,k) = \frac{\left[m^2+k^2\pi^2\right]^3}{m^2} + \frac{\beta^2k\pi}{m^2} \left(1+2n\right)\,.
\label{rayleigh_QHO}
\end{equation}
The interpretation of our simple model's prediction is the following: the linear modes have a latitudinal structure reminiscent of the quantum harmonic oscillator (QHO) wavefunctions [Eq.~(\ref{eq:QHO_eigenfunction})]. Modes are indexed by $n$, their number of zeros along the $Y$ direction, and correspond to banana cells with $n$ cyclonic-anticyclonic reversals in the $Y$ direction, hereafter referred to as an \emph{``n-banana cell''} (see Fig.~\ref{fig:banana_cells}). For fixed $m$ and $k$, a clear hierarchy is observed among these modes, all of which are more stable than the 2D RBC roll (whose threshold is recovered for $\beta\downarrow0$): their stability increases with $n$. The most unstable mode is the latitudinally symmetric \emph{0-banana cell}, with a gaussian profile in $Y$. The second most unstable mode is the anti-symmetric \emph{1-banana cell} composed of opposite cyclonicity above and below the equator. These 1- and 0-banana cells captured by the QHO approximation presented in this section are the analogs of the antisymmetric and symmetric modes computed by Roberts~\cite{robertsPTRSA68} and Busse~\cite{busseJFM70}. As $\rayleigh$ increases further, cells with an increasing index $n$ destabilize. This scenario will be illustrated, verified, and discussed further in Sec.~\ref{sec:linear}.

%
%
%
%
\section{Methods}
\label{sec:methods}
\subsection{The Equatorial $\beta$-Convection model}
\paragraph{Discretization.}In the linear limit, Eqs.~(\ref{RME}) are spatially discretized by expanding each variable in normal modes $\exp(\mri mx)$ along the periodic direction, and Chebyshev polynomials $T_k(z)$ along the bounded direction. Owing to the presence of the parity mixing operator $\mathscr{M} = Y\partial_z + \partial_Y$, both Chebyshev rational functions $SB_n(Y)$ and $TB_n(Y)$ (which are obtained as trigonometric functions mapped onto the infinite line, see e.g.~\cite{boyd, miquelJCP17}) are used along the unbounded latitudinal direction:  
\begin{subequations}
\begin{gather}
\psi(x,Y,z,t) = \sum_{m=0}^{N_x}\sum_{n=0}^{N_Y} \sum_{k=0}^{N_z} \widetilde{\psi}_{mnk}(t)\,T_k(z)\,TB_n(Y)\exp\left(\mri mx\right) + \mathrm{c.c.}\,, \\
V(x,Y,z,t) = \sum_{m=0}^{N_x}\sum_{n=1}^{N_Y} \sum_{k=0}^{N_z} \widetilde{V}_{mnk}(t)\,T_k(z)\,SB_n(Y)\exp\left(\mri mx\right) + \mathrm{c.c.}\,, \\
\Theta(x,Y,z,t) = \sum_{m=0}^{N_x}\sum_{n=0}^{N_Y} \sum_{k=0}^{N_z} \widetilde{\Theta}_{mnk}(t)\,T_k(z)\,TB_n(Y)\exp\left(\mri mx\right) + \mathrm{c.c.}\,, 
\end{gather}
\end{subequations}
where $\mathrm{c.c.}$ denotes the complex conjugate. A sparse discretization of the linear operators results from: (i) the well-known quasi-inverse technique for Chebyshev polynomials~\cite{clenshawPCPS57, julienJCP09}, and (ii) the use of two families of Chebyshev rational functions with opposite parities, as proved in~\cite{miquelJCP17}. The sparsity of the coupling matrices proves to be important from a practical point of view in both the linear stability analysis and the nonlinear time-stepping of the equations, as explained below.
\paragraph{Linear stability analysis.} In the linear limit of the model, azimuthal modes are decoupled. Thus, for a fixed azimuthal wavenumber $m$, we seek solutions whose coefficients 
are normal modes in time: \begin{equation}[\widetilde{\psi}_{mnk}(t),\widetilde{V}_{mnk}(t),\widetilde{\Theta}_{mnk}(t)]=[\widehat{\psi}_{mnk},\widehat{V}_{mnk},\widehat{\Theta}_{mnk}]\exp\left(st\right).\end{equation} The complex growth rates $s$ and the associated meridional planforms (i.e. the coefficients $\widehat{\psi}_{mnk}$ etc.) are obtained by solving sparse generalized eigenvalue problems. The parameter space is explored rapidly by obtaining the most unstable eigenvalues with Matlab's function \texttt{eigs}, an implementation of the iterative Arnoldi method. Results are periodically checked by solving for the full spectrum by means of the robust but computationally expensive QR/QZ method (Matlab's \texttt{eig}).
\paragraph{Time-stepping of the nonlinear equations.} Nonlinear dynamics are investigated with a pseudo-spectral code, implemented in Matlab and Fortran. The equations are marched in time with the classic second order IMEX Runge-Kutta scheme. For each time-step, the linear terms are treated implicitly in spectral space by solving two sparse linear systems. The nonlinear contributions are evaluated explicitly in physical space, and full dealiasing is employed in every direction. A typical truncation is $(N_x, N_Y, N_z)=(63,127,63)$, which amounts to a collocation grid of size $(192,192,96)$ as a consequence of dealiasing. Direct and inverse transforms between spectral and physical space are implemented with the FFTW library. Simulations are initialized with small amplitude random noise and are carried out until a stationary regime is identified.  Owing to fast transforms and the necessity of modest resolutions as a result of our rescaling [Eqs.~(\ref{anisotropic_scaling})], stationary regimes are usually reached within 60 core hours (ten hours on a 6-core desktop machine with Intel Sandy Bridge E architecture) using the Matlab implementation. Therefore, our model benefits from a drastically reduced computational complexity compared to DNS of the Boussinesq equations presented next.
\subsection{The Boussinesq equations in a spherical shell}
The linear stability analysis of the conduction state in the full problem is performed by means of the code QuICC~\cite{martiGGG16}, and the nonlinear dynamics are investigated with the code \emph{Rayleigh} \footnote{{h}ttps://github.com/geodynamics/Rayleigh}. These are both pseudo-spectral codes that rely on expanding the variables in spherical harmonics and Chebyshev polynomials. For small gaps $\varepsilon\downarrow 0$ (within our distinguished limit $\varepsilon\sim \Ek$), the numerical computations become involved for three reasons: (i) a large resolution is needed in $\phi$ and $\theta$ and the number of spherical harmonics scales likes $\varepsilon^{-4}$, (ii) an $\mathcal{O}(N_\theta^2)$ Legendre transform is necessary along the latitudinal direction (as opposed to an $\mathcal{O}(N_Y\log N_Y)$ Fourier transform for the reduced model), and, as a consequence, (iii) these codes are developed for parallel architectures on High Performance Computing (HPC) clusters where the cost of communication between nodes compounds to the overall algorithmic complexity. Therefore, our run with the smallest gap $\varepsilon^2= 0.01$ presented in 
Fig.~\ref{fig_rayleigh_DNS} has a resolution $ (N_\phi,N_\theta,N_r)=(4608,2304,96)$, which corresponds to 64 dealiased Chebyshev polynomials and spherical harmonics with degree less than $\ell_{\mathrm{max}}=1536$. This high resolution run necessitated 396 cores (Intel Haswell architecture) for three days before reaching a stationary regime, which amounts to approximately 30,000 core hours. 
%
%
%
%
\section{Linear Stability Analysis}
\label{sec:linear}
\begin{figure}
\includegraphics[width=0.95\textwidth]{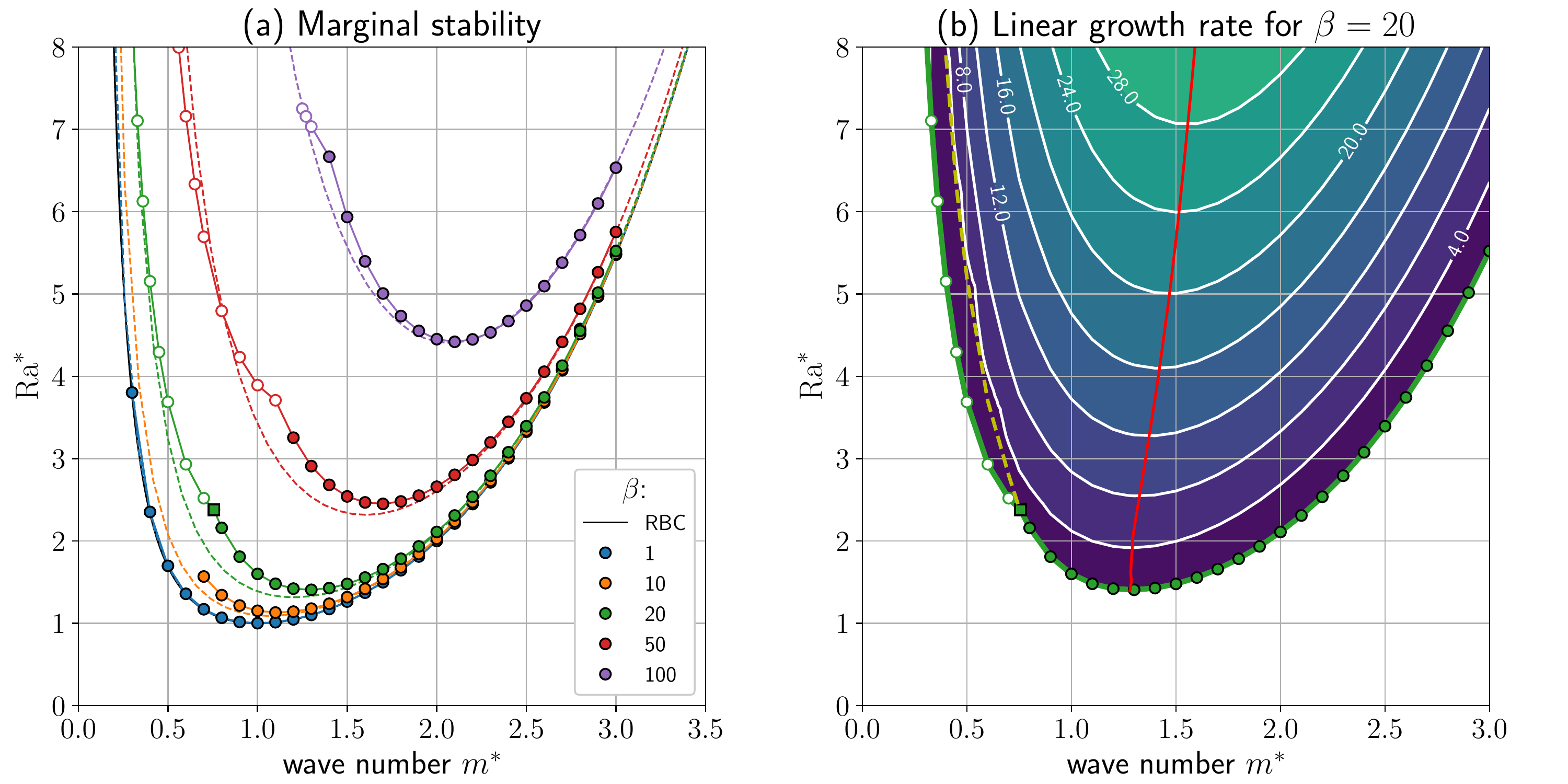}
\caption{ \label{fig_marginal_stability_EBC} (a) Marginal stability curves for the reduced model~[Eqs.~(\ref{RME})]. Filled symbols indicate steady onset ($s=0$) while open symbols indicate oscillatory onset (${\rm Re}(s)=0$, ${\rm Im}(s)\neq 0$). Dashed lines: the QHO approximation [Eq.~(\ref{rayleigh_QHO})]. (b) Isocontours of the linear growth rate $\mathrm{Re}(s)$ for the reduced model~[Eqs.~(\ref{RME})], with $\beta=20$. Green symbols indicate marginal stability $\mathrm{Re}(s)=0$. The red line marks the optimal wave number such that $\partial [\mathrm{Re}(s)] / \partial m^*=0$, for a given supercriticality. The nature of the most unstable mode changes from steady to oscillatory at the dashed yellow line. The green square indicates the Takens-Bogdanov point.
}
\end{figure}

\subsection{The \EBC model}
\subsubsection{Onset}
We first investigate the linear stability of the conduction state $\psi=V=\Theta=0$ of the \EBC model by seeking solutions of the linearized model as products of normal modes in $x$ and $t$ and meridional planforms, as detailed above. We focus exclusively on the case of $\prandtl = 1$ with stress-free and fixed temperature boundary conditions, namely $\psi = \partial_{zz}\psi =\partial_z V= \Theta = 0$ on the boundaries $z=-1,0$. The marginal curves and wave numbers are normalized by the onset Rayleigh number $27 \pi^4 / 4$ and critical wave number $\pi/\sqrt{2}$ for Rayleigh-B\'enard convection (RBC)~\ccite{chandrasekharBOOK}:
\begin{equation}
\mathrm{\rayleigh}^* = \frac{4}{27\pi^4}\rayleigh,\quad \mathrm{and}\quad m^* = \frac{\sqrt{2}}{\pi}m\,.
\end{equation}
Figure~\ref{fig_marginal_stability_EBC}(a) shows the normalized marginal stability curves $\rayleigh^*$ as a function of the normalized wave number $m^*$ for different values of the trapping parameter $\beta$. The figure shows that the conduction state in the \EBC model is more stable than in RBC, i.e. $\rayleigh^*_\beta(m) > \rarb^*(m)$ for all $m^*$. For small $\beta\approx 1$, the two curves $\rayleigh^*_\beta$ and $\rarb^*$ are almost undistinguishable. As the confinement parameter $\beta$ increases both the instability threshold $\rayleigh_c^* (\beta)$ and the corresponding critical wave number $m^*_c$ increase monotonically [Fig.~\ref{fig_marginal_stability_EBC}(a) and Fig.~\ref{fig_marginal_stability_sphere}(a)]. We note that the marginal stability curve is well-captured by Eq.~(\ref{rayleigh_QHO}), despite the crudeness of the starting point of this model [Eq.~(\ref{wild_ansatz})]. In all cases the onset is to steady convection ($s=0$) although oscillatory onset ($s=i\omega_c\ne0$) can take place for $m^*\ne m^*_c$ [Fig.~\ref{fig_marginal_stability_EBC}(a)].

We represent in Fig.~\ref{fig_marginal_stability_EBC}(b) the linear growth rate for the optimal mode as a function of the wave number $m^*$ and supercriticality $\rayleigh^*$, for constant $\beta=20$. We observe a similarity between the shape of the isocontours of the growth rate for the \EBC and RBC. The optimal wave number which corresponds to the largest growth rate for a given supercriticality (the red line on Fig.~\ref{fig_marginal_stability_EBC}b) is a slowly increasing function of $\rayleigh^*$, which is reminiscent of RBC as well. 

We close this section by commenting on the oscillatory modes observed for \EBC convection (in contrast to RBC): these oscillatory modes form a tongue of unstable modes in the $(\rayleigh^*,m^*)$ plane. A Hopf bifurcation occurs along the yellow dashed line in Fig.~\ref{fig_marginal_stability_EBC}(b). To the left of this line (small wave numbers), pairs of counter-propagating modes exist. For $\beta=20$, we find that the Takens-Bogdanov point (at the intersection of the Hopf bifurcation and the marginal stability curve) lies at the position $(m^*_{\mathrm{TB}},\mathrm{Ra}^*_{\mathrm{TB}})\approx(0.75,2.38)$. For $\prandtl=1$, we observed that $\mathrm{Ra}^*_{\mathrm{TB}}$ is always significantly larger than the onset $\rayleigh^*_c$, regardless of the value of $\beta$. Furthermore, for a given supercriticality, the growth rates of the oscillatory modes are significantly smaller than the optimal growth rate. We conclude that, in agreement with our observations, the tongue of oscillatory modes does not play any significant role in the nonlinear dynamics for $\prandtl=1$.
\subsubsection{Marginal modes}
\begin{figure}
\includegraphics[width=1.0\textwidth]{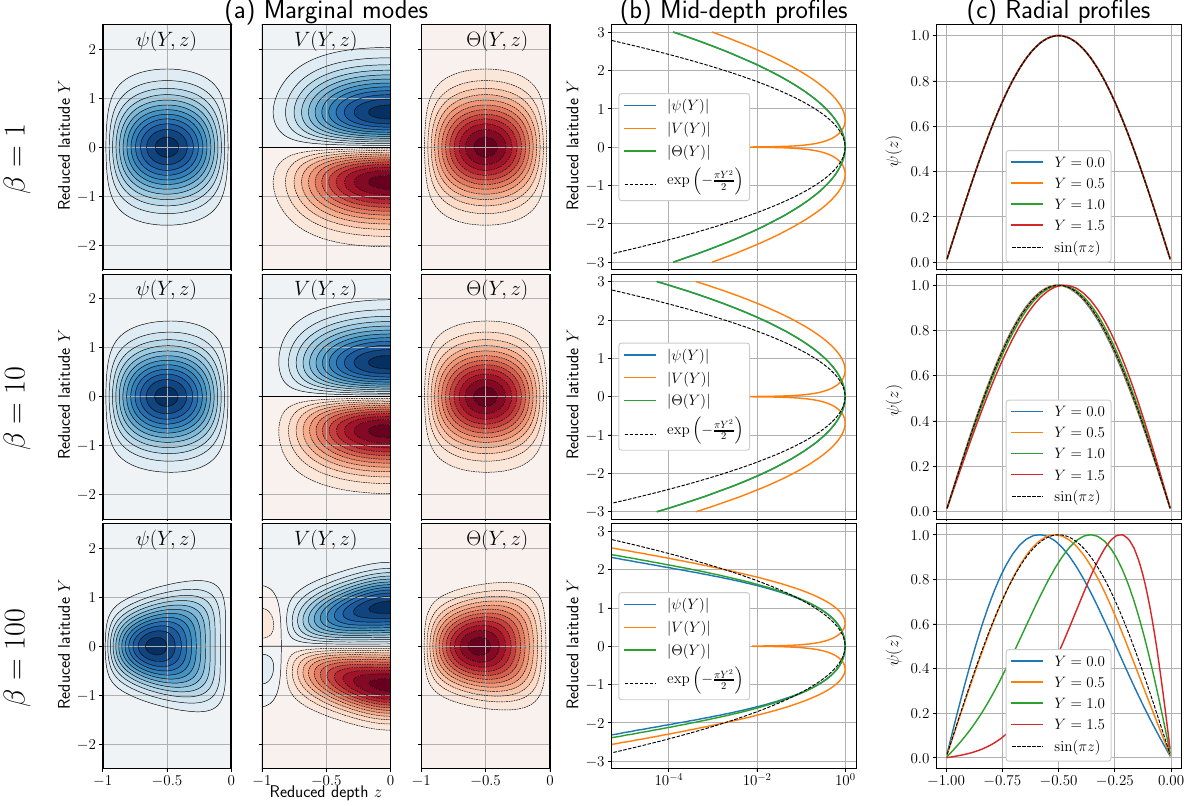}
\caption{\label{fig_marginal_modes_EBC} Marginal modes for the reduced model~[Eqs.~(\ref{RME})] when $\beta = 1$ (top row), $10$ (middle row) and $100$ (bottom row). Left column (a): meridional planforms. Center column (b): mid-depth profiles (solid lines) compared to the QHO model [Eq.~(\ref{eq:QHO_eigenfunction}), dashed line]. Right column: normalized radial profiles $\psi(z,Y)/\max_{z}(\psi(z,Y))$ at latitudes $Y=0,0.5,1,1.5$ (solid lines) compared to the ansatz~(\ref{wild_ansatz}), plotted in dashed lines. }
\end{figure}
The marginal modes for the reduced model [Eqs.~(\ref{RME})] are represented in Fig.~\ref{fig_marginal_modes_EBC} for $\beta=1$, $10$, and $100$. Meridional planforms are displayed in Fig.~\ref{fig_marginal_modes_EBC}(a). All three fields $\psi$, $V$, and $\Theta$ resemble banana cells: the fields display localization around the equator and correspond to a flow that resembles a single convective roll spanning the gap, and tapering away as the latitude increases ($|Y|\rightarrow \infty$). Perhaps expectedly, the stream function $\psi$ and the temperature profiles $\Theta$ of the marginal modes have a symmetric structure about the equator: these fields remain unchanged under the transformation $Y\leftrightarrow-Y$, regardless of the value of $\beta$. In contrast, the meridional velocity $V$ is antisymmetric about the equator. We observe that both $\psi$ and $\Theta$ are almost perfectly symmetric about mid-depth $z=-1/2$ for $\beta=1$. This symmetry degrades as $\beta$ increases: for $\beta=100$, both modes are of no particular symmetry along the radial direction. Regardless of the value of $\beta$, no particular symmetry is observed for $V$ along $z$.

We quantify the confinement of the modes along $Y$ and at mid-depth in Fig.~\ref{fig_marginal_modes_EBC}(b). We first remark that $|\psi|$ and $|\Theta|$ are very similar to each other when $\beta=1$, to the point that they cannot be distinguished in Fig.~\ref{fig_marginal_modes_EBC}(b). Both fields have a gaussian profile $\exp\left(-Y^2\right)$, which is slightly wider than the prediction (\ref{eq:QHO_eigenfunction}), whereas $V\sim Y\exp(-Y^2)$. As $\beta$ increases, the latitudinal width of the modes decreases and the fields $|\psi|$ and $|\Theta|$ become significantly distinct from one another.

We represent in Fig.~\ref{fig_marginal_modes_EBC}(c) the radial profiles of $\psi$ at different latitudes $Y$. For $\beta=1$, the stream function is almost indistinguishable from $\sin \left(\pi z\right)$ regardless of the latitude $Y$. This resemblance to 2D RBC supports our ansatz~(\ref{wild_ansatz}) for the QHO approximation, and thereby provides an explanation for the accuracy of the prediction of the threshold for instability in the \EBC model using the QHO approximation, as quantified in Fig.~\ref{fig_marginal_stability_EBC}(a). As $\beta$ increases, the profile along $z$ is no longer independent of $Y$, indicating that a separable form such as Eq.~(\ref{wild_ansatz}) is no longer an accurate approximation. The mode is no longer symmetric with respect to mid-depth. Instead, $|\psi|$ shifts towards the inner sphere at the equator and towards the outer sphere at high latitudes. This follows intuition: we recall that $\beta$ is a measure of rotation. Therefore, as rotation increases, the mode tends to straighten instead of following the weak curvature of the bounding surfaces.

\subsubsection{Secondary modes}
\begin{figure}
\includegraphics[width=\textwidth]{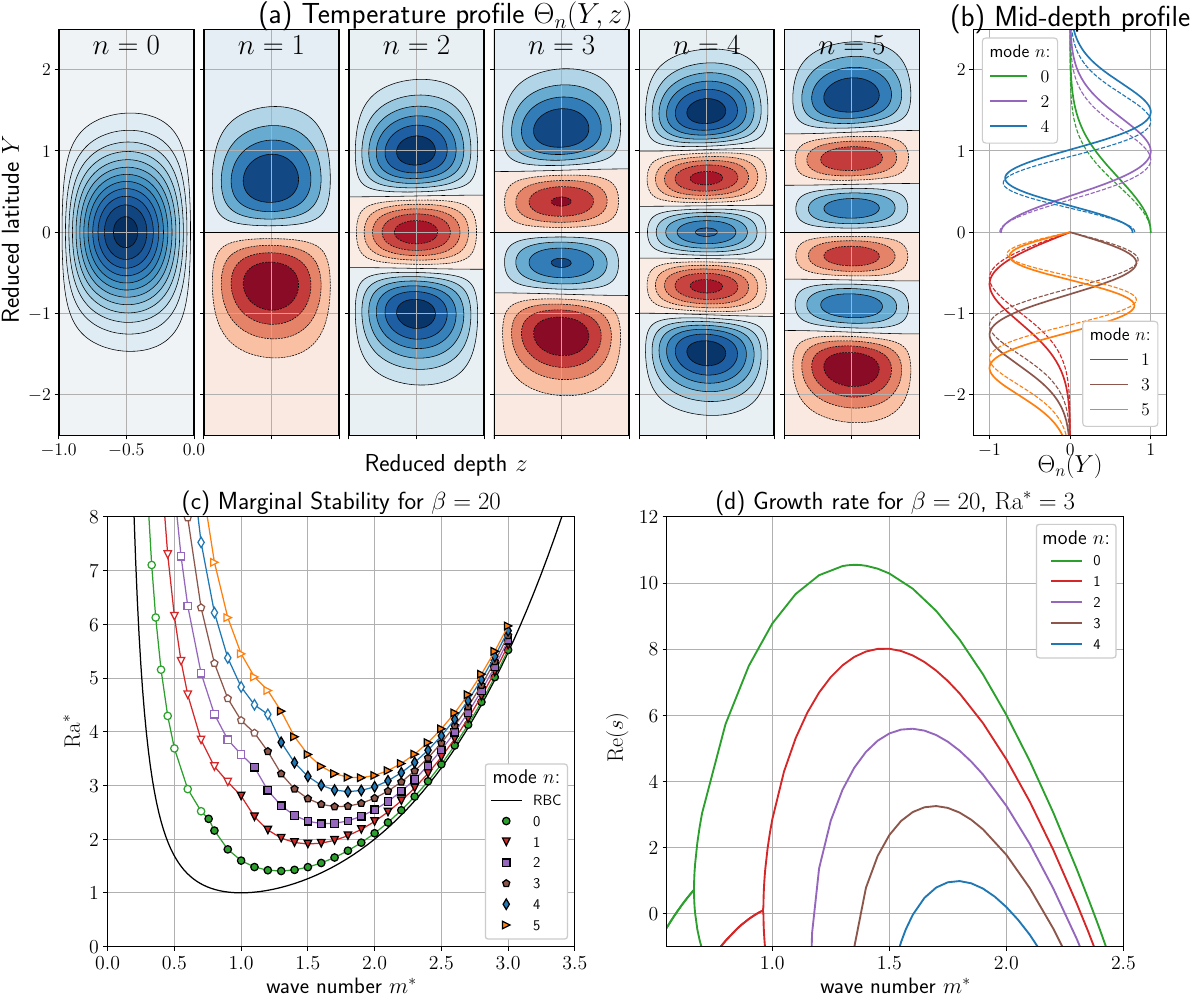}
\caption{Structure of \emph{n-banana cells} and their stability for $\beta=20$ and $0\le n\le 5$. (a) Meridional profiles of the temperature perturbations $\Theta_n(Y,z)$ at their onset. (b) Solid lines: corresponding mid-depth profiles. Dashed lines: latitudinal profiles obtained from the QHO approximation [Eq.~(\ref{eq:QHO_eigenfunction})]. (c) Marginal stability curves; filled (resp. open) symbols indicate steady (resp. oscillatory) onset. (d) Linear growth rate $\mathrm{Re}(s)$ as a function of the wave number $m^*$ for $\mathrm{Ra}^*=3$.  \label{fig_branches_EBC}}
\end{figure}
We analyze in this section the structure of the spectrum of the linearized \EBC model. We have shown above that the most unstable mode corresponds to a flow we called a \emph{0-banana cell}, i.e. a flow composed of a single convection roll along the radial direction, symmetric about and maximal at the equator, which tapers towards high latitudes. In the light of the model described in Sec.~\ref{sec:QHO_approx}, and particularly following Eq.~(\ref{eq:QHO_eigenfunction}), we expect to observe modes which are composed of an arbitrary number $n+1$ of counter-rotating cells in latitude (see Fig.~\ref{fig:banana_cells}). These cells are separated by a number $n$ of isosurfaces where the temperature perturbation vanishes. We dub these modes \emph{$n$-banana cells}, and represent them for $\beta=20$ in Fig.~\ref{fig_branches_EBC}(a,b). From Eq.~(\ref{rayleigh_QHO}), a clear hierarchy exists within these modes: their onset increases with $n$. This prediction is validated in Fig.~\ref{fig_branches_EBC}(c), where the marginal stability curves of banana cells with $0\le n \le 5$ are represented for $\beta=20$. These curves are shifted with respect to each other but share similar properties to the $n=0$ curve, already discussed. In Fig.~\ref{fig_branches_EBC}(d), we display the growth rate of these modes as a function of the azimuthal wave number $m^*$ with the supercriticality fixed at $\rayleigh^*=3$. A clear hierarchy is visible in this representation as well: as $n$ increases, the interval of unstable modes [with $\mathrm{Re}(s)>0$] shrinks, and the growth rate decreases for a given $m^*$. We relegate to Appendix~\ref{appendix:QHO} the generalization of the QHO approximation to obtain predictions for the curves presented in Fig.~\ref{fig_branches_EBC}(d).
\subsection{Comparison with the full shell} 
\begin{figure}
\includegraphics[width=\textwidth]{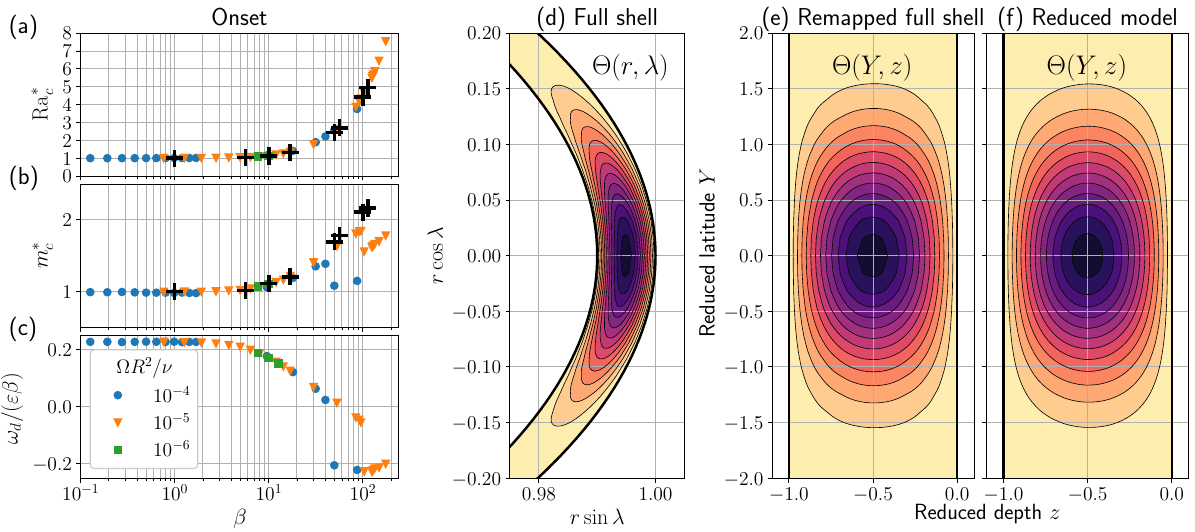}
\caption{ (a) Normalized threshold $\rayleigh^*_c$ and (b) critical wave number $m^*_c$ for both the Boussinesq equations~[Eqs.~(\ref{NSE}), color symbols] and the reduced model [Eqs.~(\ref{RME}), black crosses] as functions of $\beta$. (c) Normalized drift frequency $\omega_d/(\varepsilon\beta)$ at onset. (d,e,f) Meridional temperature profiles of the onset mode in the Boussinesq equations and in the reduced model for the parameters $\Ek=10^{-2}$, $\varepsilon=0.1$ (shell) and $\beta=10$ (reduced model). To facilitate comparison, the marginal mode of the full shell is mapped onto the reduced coordinates $(Y,z)$ in panel (e).\label{fig_marginal_stability_sphere}}
\end{figure}
The fidelity of the \EBC model is reckoned by comparing its linear properties against the linear stability results of the primitive equations~(\ref{NSE}). For full shells, the marginal modes are obtained by means of the spectral code QuICC~\ccite{martiGGG16}. For three dimensionless rotation rates $\Omega R^2 / \nu =10^{4},\,10^{5}$, and $10^{6}$, we vary the dimensionless gap $\varepsilon^2$: each full shell data point corresponds to a value $\beta = \varepsilon / E$ in the \EBC model. Onset quantities such as the threshold $\rayleigh^*_c$ and the critical wave number $m^*_c$ are plotted as functions of $\beta$ in Figs.~\ref{fig_marginal_stability_sphere}(a) and (b) for both the \EBC model and full shells. An excellent agreement between the two models is observed. However, in the shell the onset of convection is no longer steady, and all states drift slowly in the rotating frame. Figure~\ref{fig_marginal_stability_sphere}(c) shows the drift frequency $\omega_d$ normalized with $\varepsilon\beta$. A transition from retrograde drift ($\omega_d>0$)~\ccite{zhangJFM92} to the well-known prograde drift ($\omega_d<0$) is observed as $\varepsilon$ increases. A good collapse on a master curve is obtained for small to moderate $\beta<30$. We conclude that the azimuthal drift is a ${\cal O}(\varepsilon)$ effect that vanishes in shallow shells as $\varepsilon$ decreases. This weak effect is not captured by the \EBC model whose symmetry with respect to azimuthal reflection $x\leftrightarrow -x$ suppresses drift. In reality, this symmetry is always broken and weak drift is therefore to be expected. Finally, Fig.~\ref{fig_marginal_stability_sphere}(d) depicts a meridional slice of a typical marginal temperature perturbation obtained in a spherical shell in cylindrical coordinates. When mapped on the reduced coordinates $(x,Y,z)$, the temperature planform of the marginal mode in a full shell is remarkably similar to the planform of the marginal mode obtained from the reduced model with the corresponding reduced parameters [Fig.~\ref{fig_marginal_stability_sphere}(e,f)]. 
%
%
%
%
\section{Nonlinear dynamics and localized heat flux}
\label{sec:nonlinear}
\subsection{The \EBC model}
We first present the heat flux in the saturated state of the \EBC model by analyzing the behavior of the Nusselt number defined as the ratio of the azimuthally averaged convective and conductive heat fluxes:
\begin{equation}
\nusselt^\dagger(Y;\beta) = 1 - \frac{1}{L}\int_0^L\partial_z \Theta(x,Y,0)\,\mathrm{d}x\, .
\label{nusselt_EC}
\end{equation} 
\begin{figure}
\includegraphics[width=\textwidth]{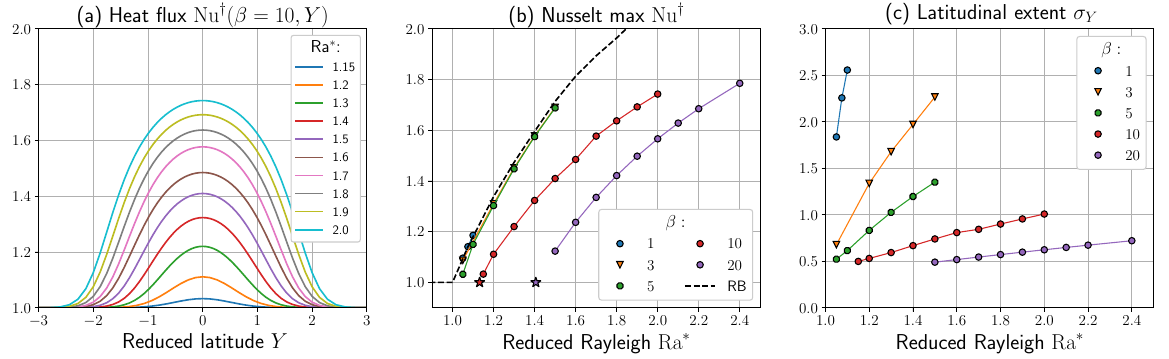}
\caption{\label{fig_saturation_reduced} Local heat transport in the saturated state of the \EBC model~[Eqs.~(\ref{RME})]. (a) Latitudinal extent of convection as measured by the heat flux $\nusselt^\dagger(Y)$ grows with increasing supercriticality when $\beta=10$. (b) Maximum Nusselt number and (c) latitudinal extent of convective motions as measured by the quantity $\stdof{Y}$ for several values of the trapping parameter $\beta$ as the supercriticality varies. The stars in panel (b) indicate instability thresholds.}
\end{figure}
This quantity, represented in Fig.~\ref{fig_saturation_reduced}, characterizes the vigor of convection and its contribution to heat transport at a given latitude $Y$. For fixed $\beta$, we observe that $\nusselt^\dagger$ peaks around, and reaches its maximum on, the equator $Y=0$. The convective heat flux decreases monotonically to its purely conductive value $N^\dagger=1$ with increasing latitude $\left|Y\right|$. Figure~\ref{fig_saturation_reduced}(a) represents $\nusselt^\dagger$ for increasing supercriticality when $\beta=10$. The figure shows that both the maximum intensity of convection and its latitudinal extent increase with supercriticality. We quantify these two features by showing in Figs.~\ref{fig_saturation_reduced}(b) and (c) the peak value ${\nusselt}^\dagger_{\mathrm{max}}(\beta)$ and the length $\sigma_Y$ computed from the second moment of ${\nusselt}^\dagger(Y;\beta)-1$,
\begin{equation}
\stdof{Y}
\equiv \left(\int_{-\infty}^{\infty} Y^2 \left[\nusselt^\dagger(Y;\beta)-1\right]\, \mathrm{d}Y \Big/\int_{-\infty}^{\infty} \left[\nusselt^\dagger(Y;\beta)-1\right]\, \mathrm{d}Y \right)^{1/2}\,,
\label{Y2_EC}
\end{equation}
both as functions of the Rayleigh number for several values of $\beta$. As $\beta$ increases at fixed supercriticality, the maximum convection amplitude decreases and so does its spatial extent in $Y$. This observation is compatible with intuition as, for a fixed gap $\varepsilon^2$, increasing $\beta$ corresponds to faster rotation (smaller Ekman number). The RBC heat flux constitutes an upper bound which, in our reduced model, is approached from below as $\beta\downarrow 0$.
 \begin{figure}[p]
\includegraphics[width= \textwidth]{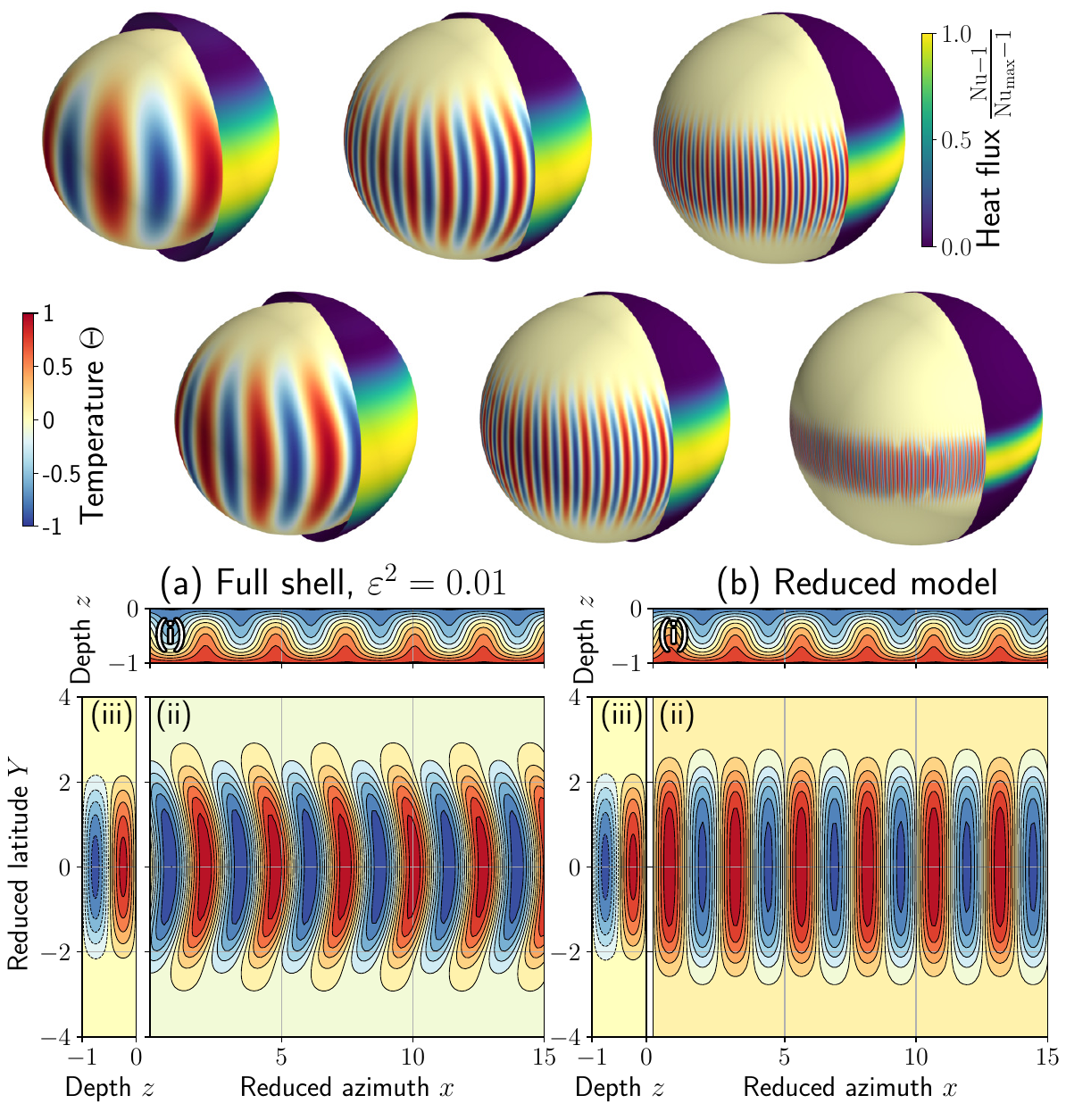}
\caption{\label{fig_rayleigh_DNS} Top panels: DNS of convection in shells of decreasing gap $\varepsilon^2 = 0.25,\, 0.16,\, 0.09,\, 0.05,\,0.03$, and $0.01$ (left to right) but constant $\beta=5$ (see text) and $\rayleigh^*=1.3$ in terms of the normalized mid-depth temperature fluctuations $\Theta/\max{|\Theta|}$ and the normalized azimuthally averaged heat flux $(\nusselt-1)/(\nusselt_\mathrm{max}-1)$ on the outer sphere. Bottom panels: temperature profiles obtained in (a) DNS with gap $\varepsilon^2=0.01$ and (b) the \EBC model, both plotted as functions of the reduced coordinates $(x,Y,z)$. In both cases we display (i) an equatorial slice at $Y=0$ and (ii) a shell slice at mid-depth $z=-0.5$ of the total temperature, along with (iii) a meridional slice of the azimuthally averaged temperature fluctuation. Azimuthal drift is visible on (a.ii) but is absent from (b.ii) as a result of the restored azimuthal symmetry in the \EBC model.}
\end{figure}
\begin{figure}[h!]
 \includegraphics[width=\textwidth]{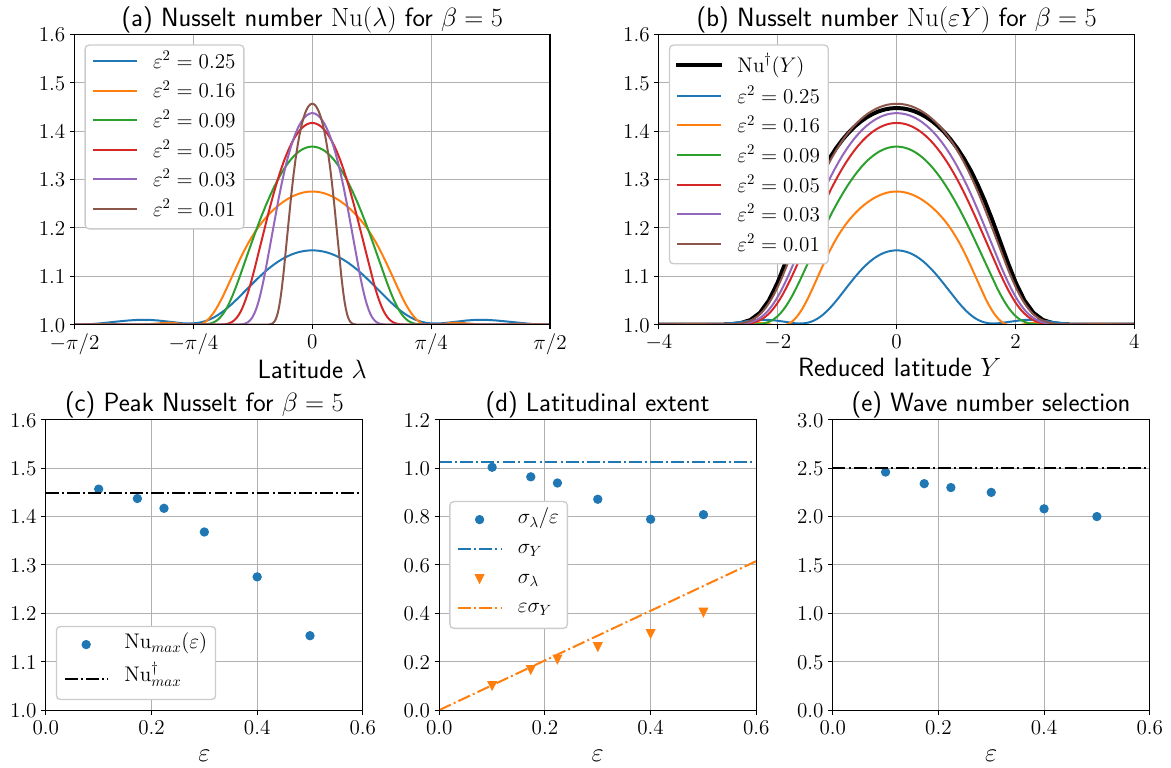}
\caption{\label{fig_SC_vs_EC}Top panels: heat flux $\nusselt$ in the six cases plotted as a function of (a) the latitude $\lambda$ and (b) the reduced latitude $Y$, for comparison with the reduced heat flux $\nusselt^\dagger(Y)$. Bottom panels: (c), (d) and (e) show $\nusselt_\mathrm{max}$, the latitudinal extent $\sigma_Y$ or $\sigma_\lambda$ [Eqs.~(\ref{Y2_EC}) and (\ref{Y2_SC})] and the scaled wave number $M=m/\varepsilon^2$ selected after saturation, all as functions of $\varepsilon$. Dashed lines indicate the predictions from the \EBC model.}
\end{figure}
\subsection{The case of full shells}
\label{section:non_linear_full_shell}
We now establish the relevance of the \EBC asymptotic regime, characterized above, for full shells. Extending the definitions (\ref{nusselt_EC}) and (\ref{Y2_EC}), we define the azimuthally-averaged dimensionless heat flux $\nusselt$ in a sphere and the associated second moment:
\begin{gather}
\nusselt(\lambda;\varepsilon,E) = 1 + \frac{1}{2\pi}\int_0^{2\pi}\frac{\partial_r \Theta(r=R,\lambda,\phi)}{\partial_r \overline{\Theta}(r=R)}\,\mathrm{d}\phi\, ,
 \\
 \stdof{\lambda}\equiv
 \left(\int_{-\infty}^{\infty} \lambda^2\, \big[\nusselt(\lambda;\varepsilon,E)-1\big]\, \mathrm{d}\lambda \bigg/\int_{-\infty}^{\infty} \big[\nusselt(\lambda;\varepsilon,E)-1\big]\, \mathrm{d}\lambda \right)^{1/2}\,.
 \label{Y2_SC}
\end{gather}
We present in Fig.~\ref{fig_rayleigh_DNS} a suite of six direct numerical simulations (DNS) of the full Navier-Stokes equations~(\ref{NSE}) in shells with gaps $\varepsilon^2 = 0.25,\, 0.16,\, 0.09,\, 0.05,\,0.03$, and $0.01$. Expecting the properties of convection to depend only on $\beta$ in the asymptotic regime, we fix $\beta =5$ and set the Ekman number of each run accordingly. The normalized Rayleigh number is set to $\rayleigh^*=1.3$ so that the corresponding Rayleigh number is $\rayleigh \approx 854.8$. We display in the top row of Fig.~\ref{fig_rayleigh_DNS} shell slices of the temperature fluctuation $\Theta$ at mid-depth together with the azimuthally averaged heat flux on the outer shell $\nusselt(\lambda)$, in pseudo-colors. In spherical coordinates, the heat flux concentrates around the equator as the gap diminishes. The temperature field obtained with the smallest gap $\varepsilon=0.01$ is represented in Fig.~\ref{fig_rayleigh_DNS}(a) as a function of the reduced coordinates $(x,Y,z)$ [see Eq.~(\ref{anisotropic_scaling})], with a view to performing a qualitative comparison with the corresponding solution to the reduced equations represented in Fig.~\ref{fig_rayleigh_DNS}(b). For both situations, we display an equatorial slice ($Y=0$) and a mid-depth shell slice ($z=-0.5$) of the full temperature profile.
These slices are complemented with meridional slices of the azimuthally averaged temperature perturbation. The \EBC model captures faithfully the azimuthal wavelength of the banana cells after saturation, as well as their latitudinal extent. Both quantities are analyzed more quantitatively below. A striking difference between the two temperature profiles in Figs.~\ref{fig_rayleigh_DNS}(a) and (b) is the bending of the banana cells in the case of the DNS. This bending results from the generation of an underlying azimuthal prograde jet at the equator of the sphere, which is itself a consequence of the breaking of the $\phi \leftrightarrow -\phi$ symmetry in Eqs.~(\ref{NSE}). As a consequence of the restored azimuthal symmetry $x\leftrightarrow -x$ in Eqs.~(\ref{RME}), no jet or bending of the convection cells is observed in the reduced model. However, the absence of the (higher order) symmetry-breaking terms in the \EBC model does not appear to impact the capacity of the model to provide accurate predictions of the heat flux at modest supercriticality, as we now demonstrate.  

A more quantitative analysis of the heat flux is provided in Figs.~\ref{fig_SC_vs_EC}(a) and (b) where this flux is plotted as a function of the latitude $\lambda$ and the reduced latitude $Y=\lambda/\varepsilon$, respectively. Figure~\ref{fig_SC_vs_EC}(b) provides evidence that, as $\varepsilon$ diminishes, the heat flux profile $\nusselt(\lambda; \varepsilon, \Ek)$ approaches an asymptotic profile $\nusselt^\dagger(Y;\beta)$ with $\beta=\varepsilon\Ek^{-1}$ and $Y=\varepsilon^{-1} \lambda$. The convergence to the asymptotic profile is documented in Figs.~\ref{fig_SC_vs_EC}(c) and (d) where $\nusselt_\mathrm{max}$ and $\sigma_\lambda$ are represented as functions of $\varepsilon$ and compared against $\nusselt^\dagger_\mathrm{max}$ and $\sigma_Y$. As $\varepsilon$ decreases both $\nusselt_\mathrm{max}$ and $\sigma_\lambda$ approach from below the corresponding asymptotic values  predicted by the \EBC model \footnote{An exception to this general trend arises for the shallowest shell considered here for which the peak heat flux may reach a value slightly above the prediction from the \EBC model. The result comes with small error bars, however, associated with the saturated heat flux obtained from both DNS and the \EBC model, and arising from numerical truncation. Owing to the prohibitive cost of DNS at such narrow gap width, we have not increased the resolution or explored even shallower shells to confirm that this very slight overshoot is a real effect.}.
\begin{table}
\caption{\label{tab:moons_features}Physical properties of the outer ice crust and the subsurface ocean of Europa and Enceladus. We provide an order of magnitude for the characteristic kinematic viscosity $\nu_{(\beta=20)}$ that would result in parameter $\beta=20$. 
}
\begin{ruledtabular}
\begin{tabular}{cccccccc}
   Moon   & Radius $R$ (km) &  Gap $\ell$ (km) & Ratio $\ell/R$ & $\varepsilon$ & $\Omega$ (rad/s) & $\nu_{(\beta= 20)}$ (m$^2$/s)  \\
   Europa (crust)    &  $1565$         & $ 15$            &  $0.0096$        & $0.098$  & $2.4\times 10^{-5}$    &  $\sim$200  \\
   Europa (ocean)    &  $1550$         & $100$            &  $0.065$        & $0.25$  & $2.4\times 10^{-5}$  &  $\sim$8000  \\
Enceladus (crust)&  $ 250$         & $25$             &  $0.1$       & $0.32$  & $5.3\times 10^{-5}$ & $\sim$1000 \\
Enceladus (ocean)&  $ 225$         & $26$             &  $0.115$       & $0.34$  & $5.3\times 10^{-5}$ & $\sim$1000
\end{tabular}
\end{ruledtabular}
\end{table}
We remark that the wave number selection is accurately captured by our reduced model as well [Fig.~\ref{fig_SC_vs_EC}(e)]. The figure shows the wave number $m$ containing the most energy in spectral space. We recall that an azimuthal mode $\exp\left(\mri M \phi\right)$ in the full sphere corresponds to the mode $\exp\left(\mri m x\right)$ with $m=M\varepsilon^2$ in the reduced geometry. Figure~\ref{fig_SC_vs_EC}(e) confirms that the wave number selected in our suite of spherical DNS approaches from below the value $m\approx 2.499$ predicted by the reduced model.

  \subsection{Applicability of the \EBC model}
  
In summary, the robustness of the predictions from the asymptotic model is quantitatively good: for instance, the error in both the latitudinal extent and the intensity of convection is within ten percent for $\varepsilon$ as big as $\sqrt{0.05} \approx0.22$. This latter value is representative of Europa's and Enceladus' subsurface oceans (see Table~\ref{tab:moons_features}), and larger than the depth of their crusts. However, for both oceans and crusts, the crux in the applicability of the model resides in our assumption that $\beta=\mathcal{O}(1)$ [Eq.~(\ref{distinguished})]. Indeed, values based on molecular diffusivities for liquid water ($\sim 10^{-6}$ m$^2$/s) are incompatible with this hypothesis, yielding a large $\beta$ for liquid oceans when estimated values of $\ell$ and rotation rates $\Omega$ typical of icy moons are considered. In contrast, evaluating $\beta$ relevant to ice crust yields equally problematic small values. In fact water-ice in planetary context is often modelled with an effective viscosity $\gtrsim 10^{11}$ m$^2$/s~\cite{pappalardoNAT98} despite its non-Newtonian character. Therefore without greater understanding of the physical rheology of both oceans and crusts, the direct applicability of the \EBC model to planets or their moons is not straightforward. Nevertheless we document estimates of the effective viscosity compatible with the asumption that $\beta=\mathcal{O}(1)$ in Table~\ref{tab:moons_features} for future reference.

\label{sec:discussion}
%
%
%
%
\section{Concluding remarks} 
\label{sec:conclusion}
We have characterized the trapping of convection columns at low latitudes in narrow rapidly rotating shells by analyzing the local convective heat flux. This trapping is faithfully described by an asymptotically reduced model valid in the distinguished regime where the Ekman number $\Ek$ is comparable to $\varepsilon$, the square root of the dimensionless gap, as confirmed by DNS of the full equations. The DNS show that as $\varepsilon$ diminishes, the flow converges to an asymptotic regime well captured by the \EBC model presented in Eqs.~(\ref{RME}). In this regime azimuthal drift is a higher order effect and so is slow. Although we cannot exclude the possibility of windy-convection in the \EBC model at much larger Rayleigh numbers analogous to that present on a plane layer~\cite{goluskinJFM14, hardenbergPRL15, guervillyGJI17}, the slow drift remains a higher order effect at the modest supercriticalities reported in this paper and therefore exerts no quantifiable effect on heat transport. Thus, the parameter $\beta=\varepsilon\Ek^{-1}$ controls both the linear stability of the conduction state and the properties of the saturated regime. Low values of $\beta$ yield weakly trapped convection, whose properties are well approximated by 2D RBC. As $\beta$ increases and trapping intensifies, the localization around the equator becomes tighter and the vigor of convection declines unless compensated by increased $\rayleigh$. The \EBC model offers predictions for narrow shells at a modest computational cost that are beyond current DNS capabilities, and has potential to be extended to MHD flows, in the spirit of existing studies in plane layer geometry~\cite{calkinsJFM15}.

In future work, the present study will be extended to larger supercriticalities (yet not so large as to trigger instability at the poles) in order to investigate the effects of spatial modulations on convection within the tropical region. Prandtl numbers different from unity will also be studied to account more rigorously for solid-state convection in ice crusts. Also required are synergistic investigations of the \EBC model in conjunction with detailed DNS studies of the primitive equations~\cite{soderlundNAT14}, as done successfully in a horizontal plane layer geometry with vertical rotation and gravity. Such an effort may lead to further developments and finer tuning of the asymptotic model.
\acknowledgements This work was supported in part by the National Science Foundation under grants DMS-1317666 (BM and KJ), NASA-NNX17AM01G (BM, NF and KJ), and DMS-1317596 (JHX and EK). NF was supported by NSF-0949446 and NSF-1550901 during the development of \emph{Rayleigh}. This work utilized the RMACC Summit supercomputer, which is supported by the National Science Foundation (awards ACI-1532235 and ACI-1532236) and is a joint effort of the University of Colorado, Boulder, and Colorado State University. Visualizations on spherical surfaces and volume renderings were obtained with the 3D visualization Python library Mayavi~\cite{MAYAVI}.
\bibliography{EC_i}
%
%
%
%
%
%
%
%
%
%
%
%
%
%
\appendix
%
%
%
%
%
%
%
%
%
%
%
%
%
%
%
\section{Derivation of the Equatorial $\beta$-Convection model}
\label{appendix:derivation}
\subsection{Convection in a shell: formulation}We start the derivation of the \EBC model by recalling the Boussinesq equations for a spherical fluid shell of gap $\ell$ and outer radius $R$, in the presence of a dimensionless background temperature profile $\overline{\Theta}(r^*)  = \overline{\Theta}_0 +  R\left(R-\ell\right) /(\ell r^*)$, with $r^*$ (resp. $r=r^*/\ell$) the dimensional (resp. dimensionless) radial coordinate:
\begin{subequations}
\begin{gather}
\prandtl^{-1} \mathrm{D}_t\boldsymbol{u} + \Ek^{-1}\boldsymbol{e}_\Omega \wedge \boldsymbol{u} = - \boldsymbol{\nabla}p  + \frac{\ell r}{R}\rayleigh\, \Theta\, \boldsymbol{e}_r + \boldsymbol{\nabla}^2 \boldsymbol{u} \,,\label{app_NS_momentum}\\ 
\boldsymbol{\nabla\cdot u} =0 \,,\label{app_NS_div}\\
\mathrm{D}_t\Theta +  u_r\partial_r \overline{\Theta} = \boldsymbol{\nabla}^2 \Theta\,,\label{app_NS_temp}
\end{gather}\label{app_NSE}
\end{subequations} 
and the definition of the Prandtl, Ekman, and Rayleigh numbers:
\begin{equation}
\prandtl = \frac{\nu} {\kappa}, \qquad \Ek = \frac{\nu} {2\Omega \ell^2},\qquad \rayleigh = \frac{\alpha g \ell^3 \delta\bar{\Theta}}{\kappa\nu}\,.
\end{equation}
The gap $\ell$ has been used to make lengths dimensionless. Upon defining the dimensionless gap $\varepsilon^2=\ell/R$, the dimensionless radius $r = r^*/\ell$ becomes
\begin{equation}
\label{app_A4}r = \frac{1}{\varepsilon^2} + z\,, 
\end{equation} where $z\in[-1,0]$ is the dimensionless depth. 
 No-slip (NS) or stress-free (SF) boundary conditions are imposed at $z=-1,\,0$:
\begin{subequations} 
\begin{align}
\mathrm{No{-}slip:}&\quad u_r = u_\phi = u_\theta = 0\,,\quad \mathrm{and}\quad \partial_z u_r=0\,,\label{NSE_velocity_BC_NS}\\
\mathrm{Stress{-}free:}&\quad u_r =0,\quad \partial_zu_\phi = \partial_zu_\theta = 0\,\quad\mathrm{and}\quad\partial_{zz}u_r=0.\label{NSE_velocity_BC_SF}
\end{align}\label{NSE_velocity_BC}
\end{subequations}
Fixed temperature or fixed flux boundary conditions are used for the temperature:
\begin{subequations}\label{NSE_temperature_BC}
\begin{align}
\mathrm{Fixed~temperature:}\quad \Theta &= 0\,, \\ 
\mathrm{Fixed~flux:}\quad \partial_r\Theta &= 0\,.
\end{align}
\end{subequations}
\subsection{Geostrophy}
We consider shallow shell geometries ($\varepsilon \downarrow 0$) and postulate rapid rotation ($\Ek\downarrow 0$), while remaining in the distinguished limit:
\begin{equation}
\beta = \varepsilon E^{-1}=\mathcal{O}(1)\,.\end{equation} 
Within this distinguished limit, it is natural to consider the dimensionless gap as the smallest characteristic scale for the flow. Therefore, we introduce along the latitudinal and longitudinal directions scales of the order of the gap. Furthermore, we allow for modulation on an intermediate scale $\mathcal{O}(\varepsilon)$, larger than the $\mathcal{O}(\varepsilon^2)$ gap but smaller than the radius $\mathcal{O}(1)$. Hence, the coordinates $\phi$ and $\lambda$ become:
\begin{equation}
\phi \rightarrow \varepsilon^2 x,\qquad \lambda \rightarrow \left(\varepsilon Y, \varepsilon^2 y\right)\, \label{eq:scaling_angles}
\end{equation}
and $\partial_\lambda = \varepsilon^{-1}\partial_Y + \varepsilon^{-2}\partial_y$. Recalling from Eq.~(\ref{app_A4}) that the radius itself is $r = \varepsilon^{-2} + z$, we illustrate on the latitudinal pressure gradient the mechanics of our change of variables:
\begin{equation}
\frac{1}{r}\partial_\lambda p = \left[ \varepsilon^{2} + \mathcal{O}\left(\varepsilon^4\right)\right] \left(\varepsilon^{-1}\partial_Y + \varepsilon^{-2}\partial_y \right) p =\partial_y p + \varepsilon \partial_Y p  +  \mathcal{O}(\varepsilon^{3})\,.
\end{equation}
We also give the leading order expression for the rotation axis $\boldsymbol{e}_\Omega$:
\begin{equation}
\boldsymbol{e}_\Omega = \boldsymbol{e}_Y + \varepsilon Y \boldsymbol{e}_z + \mathcal{O}(\varepsilon^2)\,.
\end{equation}
Armed with this information, one can readily check, upon defining $u= u_\phi$, $v = u_\lambda$ and $w = u_r$, that the governing equation (\ref{app_NS_momentum}) for the momentum and the incompressibility condition~(\ref{app_NS_div}) become:
\begin{subequations}
\begin{multline}
\prandtl^{-1}\D_t \begin{pmatrix} u \\ v \\ w \end{pmatrix}
+\Ek^{-1}\begin{pmatrix} w  \\ 0 \\- u\end{pmatrix}
+\beta Y \begin{pmatrix}  -  v \\   u \\0\end{pmatrix} 
=\\ -\Ek^{-1}\begin{pmatrix}  \partial_x p \\ \partial_y p \\\partial_z p \end{pmatrix} 
-\beta\begin{pmatrix} 0 \\ \partial_Y p \\0 \end{pmatrix}
+ \rayleigh \Theta \boldsymbol{e}_z + 
\begin{pmatrix}  \nabla^2 u 
 \\ \nabla^2 v \\\nabla^2 w
\end{pmatrix}  
+\mathcal{O}\left(\varepsilon\right)\,, \label{app_NSE_momentum_Y}
\end{multline}
and 
\begin{equation}
\Ek^{-1}\left(\partial_zw  + \partial_y v + \partial_x u\right) + \beta\,\partial_Y v = \mathcal{O}\left(\varepsilon\right)\,, \label{app_NSE_incomp_y} 
\end{equation}\label{app_NSE__expansion}
\end{subequations}
where
\begin{subequations}
\begin{gather}
\D_t = \partial_t + u\partial_x + v\partial_y + w \partial_z 
\,,\\
\nabla^2  =\partial^2_{xx} + \partial^2_{yy} + \partial^2_{zz} \,.
\end{gather}
\end{subequations}
The mass conservation equations [Eq.~(\ref{app_NSE_incomp_y})] has been multiplied through by $\Ek^{-1}$ so as to obtain a skew-adjoint operator [see Eq.~(\ref{app_geostrophic_operator}) below]. Each variable $\boldsymbol{u}$, $p$ and $\Theta$ is now expanded in a power series in $\varepsilon$:
\begin{subequations}
\begin{align}
\boldsymbol{u}(x,y,Y,z)& = \boldsymbol{u}_0 + \varepsilon \boldsymbol{u}_1 + \varepsilon^2 \boldsymbol{u}_2+\dots\,, \\
 p(x,y,Y,z)& = p_0 + \varepsilon p_1 + \varepsilon^2 p_2+\dots\,, \\
\Theta(x,y,Y,z) & = \Theta_0 + \varepsilon {\Theta}_1 + \varepsilon^2 {\Theta}_2+\dots\,.
\end{align}
\end{subequations}
Substituting these expansions into Eq.~(\ref{app_NSE}) and collecting the leading order terms at $\mathcal{O}(\Ek^{-1})$, one obtains the relation describing geostrophic balance between the Coriolis force and the pressure gradient on the one hand, complemented by an incompressibility condition on the other:
\begin{subequations}
\begin{gather}
\begin{pmatrix}   w_0 \\0 \\-u_0 \end{pmatrix} = -\begin{pmatrix}   \partial_x p_0 \\\partial_y p_0 \\\partial_z p_0 \end{pmatrix}\,,\\
\partial_xu_0 + \partial_y v_0 + \partial_z w_0 = 0\,.
\end{gather}
\end{subequations}
Upon introduction of the geostrophic operator $\mathcal{L}$, this leading order result may be formally written in the form
\begin{equation}
\label{app_geostrophic_operator}
\mathcal{L} \begin{pmatrix} \boldsymbol{u}_0 \\ p_0 \end{pmatrix} = \boldsymbol{0}\,,\quad \mathrm{with}\quad \mathcal{L} = \begin{pmatrix} 
 0   &   0   &   1 &  \partial_x \\
 0   &   0   &   0 &  \partial_y \\
-1   &   0   &   0 &  \partial_z \\
\partial_x & \partial_y & \partial_z & 0  
\end{pmatrix}\, .
\end{equation}
This operator is skew-adjoint, so that the adjoint operator satisfies $\mathcal{L}^\dagger = - \mathcal{L}$. Thus, the kernel of both the operator $\mathcal{L}$ and its adjoint $\mathcal{L}^\dagger$ is:
\begin{equation}
\mathrm{Ker}\left(\mathcal{L}\right) =\mathrm{Ker}\left(\mathcal{L}^\dagger\right)= \left\{ \begin{pmatrix}\partial_z\psi,&V,&-\partial_x \psi,&\psi\end{pmatrix}^{\mathrm{T}} \big| \:\psi,\,V\:\mathrm{arbitrary\:functions \: of\:}(x,Y,z)\right\}\,. \label{kernel}
\end{equation}
In particular, we have proved from this leading order balance the invariance of the flow along the $y$-axis (Taylor-Proudman theorem) on the fast scale $y$ for the leading order component $(u_0,v_0,w_0)$. As a consequence, the velocity $\boldsymbol{u}$ is decomposed into an equatorial stream function $\psi(x,Y,z)$ identical to the pressure, and a meridional velocity $V(x,Y,z)$, viz.
\begin{subequations}
\label{app_geostrophic_flow}
\begin{gather}
\boldsymbol{u}_0 (x,Y,z) = \partial_z\psi \,\boldsymbol{e}_x + V\, \boldsymbol{e}_y - \partial_x \psi\, \boldsymbol{e}_z\,,\\
p(x,Y,z) = \psi\,. 
\end{gather}
\end{subequations}
\subsection{Equatorial trapping}
We now derive the governing equations for the modulation on the intermediate scale $\varepsilon Y$. We proceed by collecting terms of the same order in $\varepsilon$ in Eqs.~(\ref{app_NSE__expansion}), making use of expression (\ref{app_geostrophic_flow}). The momentum equation yields
\begin{subequations}
\begin{multline}
\left(\prandtl^{-1}\partial_t-\nabla^2_\perp\right)
\begin{pmatrix}
 \partial_z \psi \\ V \\ -\partial_x \psi 
 \end{pmatrix} 
+
\prandtl^{-1}
\begin{pmatrix}
\partial_z \psi \partial_{xz}\psi - \partial_x \psi \partial_{zz} \psi \\
\partial_z \psi \partial_x V - \partial_x \psi \partial_z V \\
-\partial_z\psi \partial_{xx} \psi +\partial_x \psi \partial_{xz}\psi
\end{pmatrix}\\
+
 \beta Y \begin{pmatrix} -V \\ \partial_z \psi \\ 0 \end{pmatrix} + \beta \begin{pmatrix}  0\\ \partial_Y p \\ 0\end{pmatrix} - \rayleigh\, \Theta \, \boldsymbol{e}_z = - \beta \begin{pmatrix} w_1 \\ 0 \\ u_1 \end{pmatrix} - \beta \begin{pmatrix} \partial_x p_1 \\ \partial_y p_1 \\ \partial_z p_1 \end{pmatrix}\, .
\end{multline}
Furthermore, upon multiplication by $\beta$, incompressibility becomes
\begin{equation}
\beta \partial_Y V = - \beta \partial_x u_1 -\beta \partial_y v_1- \beta \partial_z w_1\,,
\end{equation}
\end{subequations}
so that together these equations are of the form:
\begin{equation}
\mathcal{Q}\left(\psi,V\right) = - \beta \mathcal{L} \begin{pmatrix} \boldsymbol{u}_1\\ p_1\end{pmatrix}\,. \label{eq:QL}
\end{equation}
The solvability condition for Eq.~(\ref{eq:QL}) requires $\mathcal{Q}\left(\psi,V\right)$ to be orthogonal to the kernel of the adjoint operator $\mathcal{L}^\dagger$, given above in equation~(\ref{kernel}), viz. 
\begin{equation}
\forall\:\left(\psi^\dagger,V^\dagger\right):\quad\integrate{ \begin{pmatrix} \partial_z \psi^\dagger, & V^\dagger, & -\partial_x \psi^\dagger, & \psi ^\dagger\end{pmatrix}\boldsymbol{\cdot} \mathcal{Q}\left(\psi,V\right) } 
=0\,,
\end{equation}
where $\int \mathrm{d}^3 \mathcal{V}$ represents integration over the fluid domain. Recalling the geostrophic structure of $\psi$ and $V$ [Eqs.~(\ref{app_geostrophic_flow})], the orthogonality condition takes the form:
\begin{multline}
\int\mathrm{d}^3\mathcal{V} \Bigg\{ 
\partial_z \psi^\dagger 
\bigg[ \prandtl^{-1}\big(  \partial_t+\partial_z \psi \partial_{x} - \partial_x \psi \partial_{z}\big) \partial_z \psi 
- \nabla^2_\perp \partial_z\psi
-\beta Y V
\bigg] \\
+V^\dagger 
\bigg[ \prandtl^{-1}\big(  \partial_t+\partial_z \psi \partial_{x} - \partial_x \psi \partial_{z}\big)V
-\nabla^2_\perp V+\beta Y \partial_z \psi + \beta \partial_Y \psi
 \bigg] \\
-\partial_x\psi^\dagger 
\bigg[
- \prandtl^{-1}\big( \partial_t+\partial_z \psi \partial_{x} - \partial_x \psi \partial_{z}\big) \partial_x\psi
+  \nabla^2_\perp \partial_x\psi
-\rayleigh \Theta
 \bigg] 
 +\psi^\dagger 
 \bigg[\beta \partial_Y V
  \bigg]\Bigg\} =0\,, \label{eq_bigsolva}
\end{multline}
where we recall the definition of the equatorial diffusion operator $\nabla_\perp^2 = \partial_{xx} + \partial_{zz}$. Integration by parts yields:
 \begin{multline}
 \int\mathrm{d}^3\mathcal{V} \Bigg\{ -
 \psi^\dagger 
 \bigg[
  \prandtl^{-1}\big(  \partial_t+\partial_x \psi \partial_{z} - \partial_z \psi \partial_{x}\big)\nabla^2_\perp\psi
  -\nabla^4_\perp\psi
 -\beta Y \partial_z V - \beta\partial_YV
+ \rayleigh\partial_x \Theta \bigg] \\
 +V^\dagger 
 \bigg[
   \prandtl^{-1}\big(  \partial_t+\partial_x \psi \partial_{z} - \partial_z \psi \partial_{x}\big)V
   -\nabla^2_\perp V
 +\beta Y \partial_z \psi + \beta \partial_Y \psi
  \bigg] \Bigg\} =0\,. \label{eq_bigsolvaIBP}
 \end{multline}
Finally, we obtain our reduced set of governing equations for the geostrophic fluctuations $\psi$, $V$ and $\Theta$:
\begin{subequations}\label{app_RME}
\begin{gather}\left(\prandtl^{-1}\D_t - \nabla_\perp^2\right) \nabla_\perp^2 \psi  \label{app_RM_psi}
 - \beta \left( Y \partial_z + \partial_Y\right) V = - \rayleigh\,\partial_x \Theta\,,\\ 
\left(\prandtl^{-1}\D_t - \nabla_\perp^2\right) V  + \beta \left( Y \partial_z + \partial_Y\right) \psi =0 \label{app_RM_V} \,,\\ 
 \left(\D_t - \nabla_\perp^2\right) \Theta +\partial_x \psi = 0   \, , \label{app_RM_theta}
\end{gather}
\end{subequations}
where $D_t^\perp = \partial_t + \partial_x \psi \partial_z  - \partial_z \psi \partial_x$ represents the equatorial material derivative. These equations are subject to the velocity boundary conditions (\ref{NSE_velocity_BC_NS}) which now read:
\begin{subequations}
\begin{gather}
\mathrm{No-slip:}\quad \psi = \partial_z \psi= 0\,,\quad V=0\,,\label{RME_velocity_BC_NS}\\
\mathrm{Stress-free:}\quad \psi = \partial_{zz}^2\psi = 0\,,\quad \partial_z V  =0 \label{RME_velocity_BC_SF}
\end{gather}
\end{subequations}
and the temperature boundary conditions (\ref{NSE_temperature_BC}) which read $\Theta=0$ for fixed temperature, or $\partial_z \Theta = 0 $ for fixed flux.
%
%
%
%
%
%
%
%
%
\section{Growth rate prediction from the QHO model: general case}
\label{appendix:QHO}
Starting from the linearized Eqs.~(\ref{RME})
\begin{subequations}
\begin{gather}\left(\prandtl^{-1}\partial_t- \nabla_\perp^2 \right)  \nabla_\perp^2 \psi 
=  \beta\mathscr{M} V  - \rayleigh\,\partial_x \Theta  \,, \label{RM_psi_linear}
 \\
 \left(\prandtl^{-1}\partial_t - \nabla_\perp^2\right) V = - \beta\mathscr{M} \psi\label{RM_V_linear}\,,
  \\ 
 \left(\partial_t -\nabla_\perp^2\right) \Theta=-\partial_x \psi  \,,      \label{RM_theta_linear}
 \end{gather}
\end{subequations}
we obtain:
\begin{equation}
\label{B2}
\left(\prandtl^{-1}\partial_t- \nabla_\perp^2 \right)^2\left(\partial_t- \nabla_\perp^2 \right)  \nabla_\perp^2 \psi 
=  -\beta^2\mathscr{M}^2 \left(\partial_t- \nabla_\perp^2 \right)\psi  +\rayleigh \left(\prandtl^{-1}\partial_t- \nabla_\perp^2 \right) \partial_{xx} \psi\, .
\end{equation}
We assume that the solution can be approximated by an exponentially growing counterpart of Eq.~(\ref{wild_ansatz}):
\begin{equation}
\label{exp_growing_wild_ansatz}
\psi\approxsol \approx \sin(k\pi z)\exp(\mri m x+st)\Psi (Y)\,,
\end{equation}
where $k$ is an integer. Defining $\kperp = m^2 + k^2 \pi^2$, and evaluating Eq.~(\ref{B2}) using the ansatz~(\ref{exp_growing_wild_ansatz}) along the line $z=-1/2$ where the $z$ derivative vanishes, it follows that:
 \begin{multline}
  \beta^2\left(-\partial_{YY} + k^2 \pi^2 Y^2\right)\left(s+\kperp \right)\Psi =  \rayleigh \left(\prandtl^{-1}s+\kperp \right) m^2 \Psi\\-\left(\prandtl^{-1}s+ \kperp \right)^2\left(s+\kperp \right)  \kperp \Psi \,.
 \end{multline}
 \begin{figure}
 \includegraphics[width=0.6\textwidth]{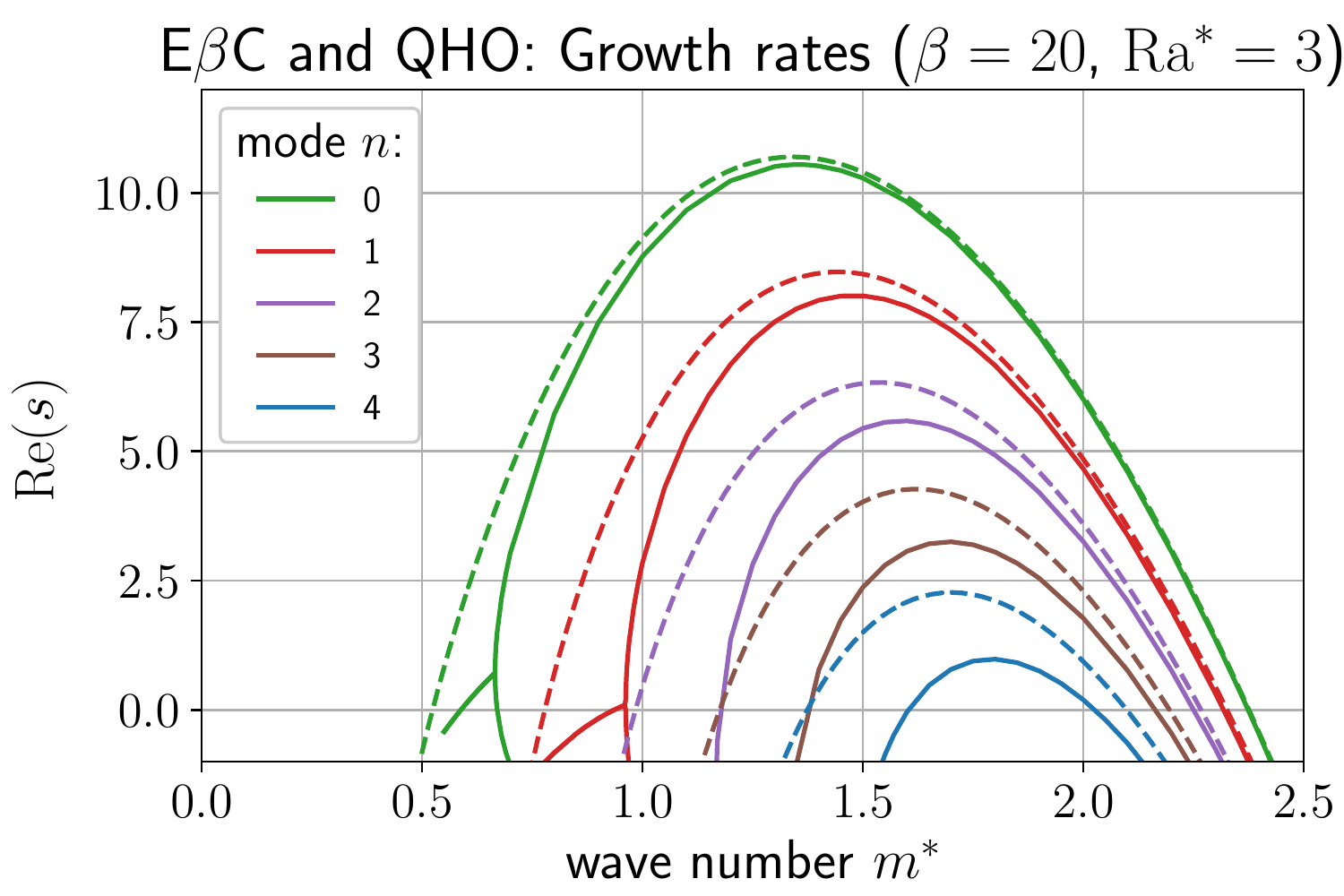}
 \caption{\label{fig:EBC_vs_QHO} Comparison of the growth rates of the first five Y-modes obtained with the \EBC model (plain lines, taken from Fig.~\ref{fig_branches_EBC}b)  and the corresponding QHO approximation [see Eq.~(\ref{real_growthrate_QHO}) with $k=1$] plotted as dashed lines.}
 \end{figure}
 Recalling the eigenvalues of the operator $\left(-\partial_{YY} + k^2 \pi^2 Y^2\right)$ from Eq.~(\ref{QHO_ev}), one obtains a cubic equation for $s$:
  \begin{equation}
   k\pi\beta^2\left(1+2n\right)\left(s+\kperp \right) =  \rayleigh \left(\prandtl^{-1}s+\kperp \right) m^2 -\left(\prandtl^{-1}s+ \kperp \right)^2\left(s+\kperp \right)  \kperp  \,.
  \end{equation}
  The case $\prandtl=1$ yields a (stable) viscous solution $s=-\kperp$ and the two roots of the quadratic equation:
\begin{equation}
k\pi\beta^2\left(1+2n\right) =  \rayleigh\,  m^2 -\left(s+ \kperp \right)^2  \kperp 
\end{equation}
given by:
\begin{equation}
s = - \kperp + \sqrt{\frac{\rayleigh\, m^2 - k\pi\beta^2\left(1+2n\right)}{\kperp}}\,.
\label{real_growthrate_QHO} 
\end{equation}
The growth rate becomes complex for
\begin{equation}
 m<\beta\sqrt{\frac{k\pi(1+2n)}{\rayleigh}}\,,
 \quad 
 \mathrm{or}
  \quad 
  m^*<\beta\sqrt{\frac{8k(1+2n)}{27\pi^5 \rayleigh^*}}\,,
 \end{equation}
but with a negative real part:
\begin{equation}
s = - \kperp + \mri\,\sqrt{\frac{-\rayleigh\, m^2 + k\pi\beta^2\left(1+2n\right)}{\kperp}}\,.    
\end{equation}
It follows that no Hopf bifurcations are present in the QHO approximation, in contrast to the \EBC model. Nonetheless, Fig.~\ref{fig:EBC_vs_QHO} presents additional evidence for the fidelity of the QHO approximation. For a given supercriticality $\rayleigh = 3$ and trapping parameter $\beta$, the figure shows the growth rates of the first five modes in the $Y$ direction obtained from the \EBC model (documented in Fig.~\ref{fig_branches_EBC}) as functions of the horizontal or azimuthal wavenumber $m^*$. These growth rates are confronted with the corresponding QHO prediction given in Eq.~(\ref{real_growthrate_QHO}), for $k=1$.

For the fundamental mode $n=0$, the agreement is excellent for $m^*>1.5$, and the QHO approximation captures well the wave number and growth rate of the optimal mode within the $m^*\in[1,1.5]$ range. However, the accuracy of the QHO approximation degrades as $m^*$ diminishes. In particular, the Hopf bifurcation for small $m^*$ is not captured. Further, fidelity of the QHO model degrades for modes with $n>0$ although large wave numbers $m^*$ continue to be described more faithfully than small wave numbers.
\end{document}